\documentclass[%
 reprint,
 amsmath,amssymb,
 aps,pra,
]{revtex4-2}
\usepackage{chemformula}
\usepackage{graphicx}
\usepackage{dcolumn}
\usepackage{bm}
\usepackage{amsmath}
\usepackage{amssymb}
\usepackage{accents}
\graphicspath{{Figures/}}
\usepackage[colorlinks]{hyperref}
\usepackage{times}
\usepackage{color}
\usepackage{float}

\definecolor{orange}{rgb}{1,0.5,0}
\definecolor{teal}{RGB}{32,178,170}

\newcommand{\sakc}[1]{ {\color{black} #1} }

\newcommand{\varg}{ \text{\it{g}} }

\newcommand{\DthT}{\tbar{D}_{\rm th}}
\newcommand{\ESS}{\tbar{E}_{\rm ss}}
\newcommand{\PSS}{\tbar{P}_{\rm ss}}
\newcommand{\DSS}{\tbar{D}_{\rm ss}}

\newcommand\tbar[1]{\accentset{\rule{.4em}{1.25pt}}{#1}}

\begin{document}

\preprint{APS/123-QED}

\title{Efficient computation of coherent multimode instabilities in lasers using a spectral approach}



\author{Sara Kacmoli*}
\email{skacmoli@princeton.edu}
\author{Saeed A. Khan*}
\email{saeedk@princeton.edu}
\author{Claire F. Gmachl}
\author{Hakan E. T\"ureci}

\affiliation{Department of Electrical and Computer Engineering, Princeton University, Princeton, NJ 08544 USA \\ *These authors contributed equally to this publication.}

\date{\today}

\begin{abstract}
Coherent multimode instabilities are responsible for several phenomena of recent interest in semiconductor lasers, such as the generation of frequency combs and ultrashort pulses. These techonologies have proven disruptive in optical telecommunications and spectroscopy applications. While the standard Maxwell-Bloch equations encompass such complex lasing phenomena, their integration is computationally expensive and offers limited analytical insight. In this paper, we demonstrate an efficient spectral approach to the simulation of multimode instabilities via a quantitative analysis of \sakc{the instability of single-frequency lasing in ring lasers, referred to as the Lorenz-Haken (LH) instability or the Risken-Nummedal-Graham-Haken (RNGH) instability in distinct parameter regimes}. Our approach, referred to as CFTD, uses generally non-Hermitian Constant Flux modes to obtain projected Time Domain equations. CFTD provides excellent agreement with finite-difference integration of the Maxwell-Bloch equations across a wide range of parameters in regimes of non-stationary inversion, including frequency comb formation and spatiotemporal chaos. We also develop a modal linear stability analysis using CFTD to efficiently predict multimode instabilities in lasers. The combination of numerical accuracy, speedup, and semi-analytic insight across a variety of dynamical regimes make the CFTD approach ideal to analyze multimode instabilities in lasers, especially in more complex geometries or coupled laser arrays.

\end{abstract}

                              
\maketitle


\section{Introduction}

Lasers are complex dynamical systems the operation of which is enabled by the interaction between the field of an often-multimode cavity and a gain medium. For a considerable part of the history of lasers, experimental efforts have been focused on improving stability, monochromaticity, power performance and beam quality. This is largely due to specific applications where these attributes are desirable, such as fiber optical telecommunication, materials processing, and precision spectroscopy. Even in such high-performance regimes, nonlinear, multimode effects emerge via mode competition and spatial hole burning, which must be understood and suppressed~\cite{ge_enhancement_2014, aung_threshold_2015} to maximize laser power and efficiency. More recently, however, rather than an undesired effect, coherent multimode lasing phenomena have been the focus of many studies~\cite{wang_coherent_2007, lugiato_self-pulsing_2019, silvestri_coherent_2020, wang_harmonic_2020, vukovic_numerical_2020, piccardo_laser_2021}.  Such phenomena include ultrashort (sub-ps) pulse formation~\cite{Tschler2021, Mirian2021} useful for precision machining~\cite{Sugioka2017} and probing of ultrafast processes~\cite{Torre2004, Huber2001}. Active mode locking is another example traditionally linked with the generation of ultrashort pulses~\cite{Morgner:99, Kim:16}; recent efforts have focused on achieving mode locking in lasers with fast gain recovery, such as quantum cascade lasers~\cite{Revin2016, Hillbrand2020}. Furthermore, soliton generation has, in recent years, proven to be a disruptive technology via passive dissipative Kerr combs~\cite{Kippenberg2018, Pfeiffer:17} as well as active platforms~\cite{mengDissipativeKerrSolitons2022a}. Optical frequency combs – whether or not characterized by ultrashort pulses – are also a prime example of a technology hinging on coherent instabilities that has been impactful in applications of spectroscopy~\cite{review_spectroscopy}, metrology~\cite{review_metrology, space_metrology}, optical clocks~\cite{Papp:14} and optical telecommunication~\cite{Gaeta2019}. 

In contrast to standard lasing regimes marked by stationary population inversion in the gain medium, coherent multimode instabilities present in the examples above are mediated by nontrivial dynamics of the lasing medium itself, referred to as \sakc{\textit{population pulsations}} in prior work~\cite{mansuripur_single-mode_2016}. \sakc{One of the earliest predicted examples of such dynamics are the instabilities of single-frequency lasing in ring lasers, often classed into Lorenz-Haken (LH, or `single-mode' ) and Risken-Nummedal-Graham-Haken (RNGH, or `multi-mode') instabilities~\cite{risken_selfpulsing_1968, graham_quantum_1968}. Either instability emerges past a \textit{second} threshold where the gain overcomes loss for additional modes, leading to steady-state multi-frequency emission. This second threshold was predicted to be at high pump powers, rendering experimental verification elusive~\cite{Hillman:85, pessina_experimental_1997, Roldn1998, voigt_experimental_2004}. However, in more recent years, low-threshold multimode instabilities have been investigated in ring quantum cascade lasers~\cite{mid_ir_qcl_comb, Piccardo2020, Jaidl:21}, as well as alternate geometries including Fabry-P\'erot lasers~\cite{mansuripur_single-mode_2016, vukovic_low-threshold_2017}, and the instability mechanism at play has been subject of much work. }


The development of simulation tools to capture \sakc{such instabilities} is thus very timely. While the semi-classical Maxwell-Bloch equations (MBEs) have provided the standard description of spatio-temporal lasing dynamics across the aforementioned single and multi-mode lasing regimes, the powerful simulation methods that have been developed to simulate MBEs are computationally expensive~\cite{demeter_solving_2013, wang_comparison_2013}, and provide only limited analytic insight into these lasing phenomena. As a result, alternative simulation methods to more efficiently analyze lasing phenomena in nonlinear dynamical regimes are highly desirable. Such schemes are also particularly relevant with increasing interest in lasers of more complex geometries or coupled laser arrays~\cite{kirch_55_2015, liu_coupled_2015, zhou_high-power_2019, Liu2022, kacmoli2023photonic}, for which simulation complexity will further increase.

In this paper, we develop and \sakc{validate} a spectral approach to simulating complex multimode laser instabilities to address this need. \sakc{While spectral or modal approaches have been used previously to analyze laser instabilities~\cite{valcarcel_modal_2003}, they use lossless, closed cavity modes as a basis, and often consider restricted parameter regimes to reduce the dimensionality of the MBEs. Our approach differs in these two keys aspects.} First, our spectral basis of choice is spanned by cold modes defined by the \textit{lossy} laser cavity: the so-called constant-flux (CF) modes of the laser~\cite{ge_2010}, which satisfy a generally non-Hermitian boundary value problem with associated complex eigenvalues. For cavities where the internal loss dominates over loss due to openness and radiation, \sakc{such as the original RNGH and LH models we compare to}, the modal theory can be cast in terms of standard closed-cavity modes with complex eigenvalues denoting intrinsic loss. \sakc{Secondly, and more importantly, we analyze the full MBEs, placing no constraints on the gain medium dynamics. This requires us to use the spectral basis to perform a spatial projection of the spatiotemporal inversion field, casting the modal theory in terms of a general \textit{time-dependent} inversion \textit{matrix}. This non-standard representation ultimately proves} crucial to correctly capturing the dynamics of the inversion field and thus of instabilities mediated by population pulsations. 

The resulting coupled-mode theory, referred to as CFTD in our earlier work~\cite{malik_spectral_2015}, has previously been applied to specific situations where only brief violations of stationary inversion occur, such as synchronization. \sakc{However, in this paper we test and demonstrate the utility of CFTD across a wide range of regimes in ring lasers characterized by nontrivial dynamics of the gain medium, encompassing both LH and RNGH instabilities, below, past, and far above the so-called second threshold. This spans dynamical phenomena ranging from decaying spatiotemporal oscillations around a single-mode lasing state, to stable broadband frequency combs, and even chaotic dynamics.} Crucially, we provide a thorough benchmarking study of CFTD across these regimes using at least one of two standard methods: finite difference time domain (FDTD) integration of multimode lasing dynamics and a split-step Runge-Kutta method (SSRK) tailored to ring lasers. We find excellent qualitative agreement across the broad range of considered regimes, with very good \textit{quantitative} agreement in specific regimes that we identify. Furthermore, the agreement is obtained via CFTD simulations several orders of magnitude faster than FDTD or SSRK simulations. 

\sakc{Finally, the CFTD approach goes beyond providing an efficient numerical tool: it also forms the foundation for a modal linear stability analysis (LSA) that we show can be used to predict multimode instabilities. The modal LSA explicitly describes the instability of \textit{discrete} modes coupled via inversion matrix elements describing gain medium dynamics. This is in contrast to more standard perturbative approaches to ring lasers, which analyze the instability of a continuous perturbation of the spatiotemporal PDE~\cite{risken_selfpulsing_1968}. Not only does the modal LSA typically agree with the spatiotemporal (ST) LSA, it can at times provide additional information. In particular, the modal LSA derived from CFTD is natively aware of the laser cavity mode spectrum, unlike the ST LSA. As a result, we find it can correctly predict the absence of instabilities in parameter regimes when a newly generated frequency does not coincide with a cavity mode, in contrast to the ST LSA.} The numerical accuracy and efficiency of CFTD simulations, together with predictive capabilities provided by the modal LSA, make it ideal to study multimode instabilities in more complex, coupled laser geometries.

This paper is organized as follows. In Sec.~\ref{sec:mbe} we \sakc{recount the standard description of lasing via MBEs, including their form within the slowly-varying envelope approximation typically employed in the analysis of ring laser instabilities.} In Sec.~\ref{sec:cftd}, we present the \sakc{CFTD approach} starting from the general MBEs, and obtain the set of time-dependent ordinary differential equations (ODEs) that constitute the CFTD equations. Sec.~\ref{sec:singleModeLasing} considers the single-mode lasing regime below the multifrequency instability threshold, applied in particular to spatially non-trivial initial field distributions involving multiple spatial modes. In Sec.~\ref{sec:rngh}, \sakc{we consider dynamics above the threshold of the ring laser instability leading to stable frequency comb formation, as a function of laser loss parameters.} In Sec.~\ref{sec:chaos} we investigate the CFTD method in a parameter space that leads to chaotic behavior, and finally in Sec.~\ref{sec:speedup} we quantify the simulation time improvement that our method has over \sakc{spatiotemporal} schemes.

\section{Maxwell-Bloch equations for ring lasers}
\label{sec:mbe}

\begin{figure}[t]
    \centering
    \includegraphics[scale=1.0]{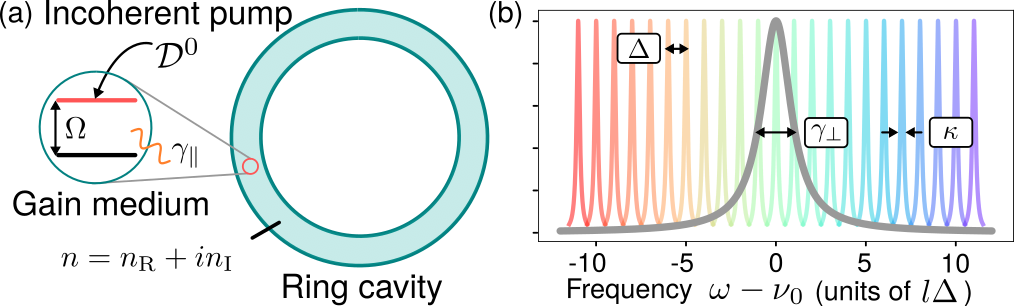}
    \caption[]{(a) Schematic of multimode ring laser under incoherent pumping. (b) Longitudinal mode structure of the ring cavity and gain medium response curve (grey). Here $N=20$ cold cavity modes are depicted. }
    \label{fig:schematic}
\end{figure}

Lasing dynamics of a variety of lasers have been very successfully described by MBEs for the electric field inside the laser cavity $\mathcal{E}(\mathbf{r},t)$ coupled to the polarization $\mathcal{P}(\mathbf{r},t)$ and inversion $\mathcal{D}(\mathbf{r},t)$ of the gain medium,
\begin{subequations}
\begin{gather}
    {\nabla}^2 \mathcal{E}-\frac{n^2}{c^2}\ddot{\mathcal{E}} ={\mu}_0\ddot{\mathcal{P}}
    \label{eq:MBE} \\
    \dot{\mathcal{P}} = (-i\Omega -{\gamma_\perp})\mathcal{P} - i\frac{\varg^2}{\hbar}\mathcal{E}\mathcal{D}
    \label{eq:MBP} \\
    \dot{\mathcal{D}} = -\gamma_\parallel(\mathcal{D} - \mathcal{D}^0) + i\frac{2}{\hbar}(\mathcal{E}\mathcal{P}^* - \mathcal{E}^*\mathcal{P})
    \label{eq:MBD}
\end{gather}
\end{subequations}
where $c = \frac{1}{\sqrt{\mu_0\epsilon_0}}$ is the speed of light in vacuum. Mathematically, Eqs.~(\ref{eq:MBE})-(\ref{eq:MBD}) most generally describe a set of coupled partial differential equations (PDEs) for arbitrary lasers. They must be accompanied by boundary conditions, here set by the geometry of the cavity confining the electric field. In this paper, we will consider the specific case of a ring laser (radius $R$ and length $L =2\pi R$), depicted schematically in Fig.~\ref{fig:schematic}~(a), for which the boundary conditions must be periodic. This enables a simplification of the MBEs to a set of scalar PDEs, defined along the ring laser arc length coordinate, which we label $x$. Then, $n$ is the effective refractive index of the cavity medium, which can be complex to incorporate cavity loss; in particular we write it as $n = n_{\rm R} + in_{\rm I}$. Phenomenological parameters $\gamma_\parallel$ and $\gamma_\perp$ represent the population and polarization decay rate, respectively, as shown in Fig.~\ref{fig:schematic}. $\Omega$ is the center frequency of the gain curve, $\varg$ is the dipole moment, and $\mathcal{D}^0$ describes incoherent pumping threshold necessary to achieve population inversion and lasing. For simplicity, we consider a spatially homogeneous pump, although the spectral approach we employ can also account for nontrivial spatial pump profiles.

We begin by considering the standard approach to Eqs.~(\ref{eq:MBE})-(\ref{eq:MBD}), namely employing the slowly-varying envelope approximation~\cite{agrawal_nonlinear_2013} to reduce the order of derivatives, reducing the wave equation for electric field evolution to an advection equation (or, in nonlinear optical media, the nonlinear Schr\"odinger's equation).


\subsection{Slowly-varying envelope approximation}

Within the standard slowly-varying envelope approximation, the form of the MBEs is simplified by explicitly extracting the spatiotemporal dependence at the frequency set by the atomic transition frequency. For ring lasers, this takes the form:
\begin{subequations}
\begin{align}
    \mathcal{E}(x,t) &= E_c\cdot\tbar{E}(x,t) \frac{1}{\sqrt{L}} e^{i (n_{\rm R} \Omega /c) x} e^{-i\Omega t} \label{eq:ansSVE} \\
    \mathcal{P}(x,t) &= P_c\cdot\tbar{P}(x,t) \frac{1}{\sqrt{L}} e^{i (n_{\rm R} \Omega /c) x} e^{-i\Omega t} \label{eq:ansSVP} \\
    \mathcal{D}(x,t) &= D_c\cdot \tbar{D}(x,t) \label{eq:ansSVD}
\end{align}
\end{subequations}
where we have also extracted dimensionful factors of the physical fields for convenience:
\begin{align}
    E_c = \frac{\hbar\sqrt{L{\gamma}_{\perp} {\gamma}_{\parallel}}}{2\varg}, 
    P_c = \epsilon_0 E_c,
    D_c = \frac{\hbar\epsilon_{0}\gamma_\perp}{\varg^2}.
    \label{eq:scalingF}
\end{align}

$\tbar{E}(x,t)$, $\tbar{P}(x,t)$ then describe the envelopes of the total electric and polarization fields respectively, which typically evolve at frequencies much slower than the large atomic transition frequency $\Omega$ that has been explicitly extracted. Substituting Eq.~(\ref{eq:ansSVE})-(\ref{eq:ansSVD}) into Eqs.~(\ref{eq:MBE})-(\ref{eq:MBD}) and dropping terms proportional to the second-order time-derivative of the envelope fields defines the slowly-varying envelope approximation. The final MBEs under the slowly-varying envelope approximation take the form of an advection equation for the electric field (instead of a second-order wave equation), coupled to ODEs for the polarization and inversion fields:
\begin{subequations}
\begin{align}
    \dot{\tbar{E}} &= -\frac{1}{n}\partial_x \tbar{E} - \tbar{\kappa} \tbar{E} + i\frac{\tbar{\Omega}}{2n_{\rm R}^2} \tbar{P} \label{eq:SVE} \\
    \dot{\tbar{P}} &= -\tbar{\gamma}_{\perp} \tbar{P} - i\tbar{\gamma}_{\perp} \tbar{E}\tbar{D} \label{eq:SVP} \\
    \dot{\tbar{D}} &= -\tbar{\gamma}_{\parallel} (\tbar{D}-\tbar{D}^0) + i\frac{\tbar{\gamma}_{\parallel}}{2} \left( \tbar{E}\tbar{P}^* - \tbar{E}^*\tbar{P} \right) \label{eq:SVD}
\end{align}
\end{subequations}
where we have also introduced the dimensionless space, time, and frequency scales:
\begin{align}
    \tbar{x} = \frac{x}{L},~\tbar{t} = \frac{t}{L/c},~(\tbar{\Omega},\tbar{\gamma}_{\parallel},\tbar{\gamma}_{\perp},\tbar{\kappa} ) = \frac{L}{c}(\Omega,\gamma_{\parallel},\gamma_{\perp},\kappa)
    \label{eq:scalingV}
\end{align}
and where $\tbar{D}^0 = \frac{D^0}{D_c}$, while $\tbar{\kappa} = \frac{n_{\rm I}}{n_{\rm R}}\tbar{\Omega}$ describes the cavity loss rate proportional to the imaginary part of the refractive index. 

Having introduced the various scaling transformations via Eqs.~(\ref{eq:scalingF}),~(\ref{eq:scalingV}), we will now drop the $\tbar{(\cdot)}$ notation in the remainder of this paper, for clarity of the presentation. All quantities from here on are therefore to be understood as dimensionless, unless otherwise noted.


\renewcommand\tbar[1]{{#1}}

Eqs.~(\ref{eq:SVE})-(\ref{eq:SVD}) and their associated periodic boundary conditions still describe a set of PDEs, and thus must be solved using a spatio-temporal integration scheme such as a FDTD method. While well-established, such schemes are computationally expensive and scale unfavourably with system size. In the following sections, we develop an efficient spectral approach to capturing \textit{dynamics} of complex multimode lasers described by the MBEs, and then benchmark this approach against more standard numerical techniques for simulating Eqs.~(\ref{eq:SVE})-(\ref{eq:SVD}).

\section{Multimode CFTD approach for laser dynamics}
\label{sec:cftd}

Any spectral approach to analyzing laser dynamics uses a spatial basis to project Eqs.~(\ref{eq:MBE})-(\ref{eq:MBD}), thereby yielding a set of \textit{ordinary} differential equations. The distinguishing feature of the CFTD spectral approach, introduced in Ref.~\cite{malik_spectral_2015}, is the use of a general set of non-Hermitian modes as a basis: the constant-flux (CF) modes. In a multi-dimensional domain $\mathcal{R}$, the CF modes $\{\varphi_m(\tbar{\bm{r}})\}$ are solutions to the generalized eigenproblem (in normalized units) $\tbar{\nabla}^2 \varphi_m(\tbar{\bm{r}}) = -n^2(\tbar{\bm{r}}) \tbar{\omega}_m^2\varphi_m(\tbar{\bm{r}})$, with complex eigenfrequencies $\{\tbar{\omega}_m\}$. Crucially, the modes obey outgoing boundary conditions past the boundary of the spatial domain, $\partial\mathcal{R}$, correctly accounting for losses due to the fields leaving the laser cavity. The non-Hermitian CF basis has successfully provided the foundation for Steady-state Ab Initio Lasing Theory (SALT)~\cite{ge_2010}.



While the CFTD method has been developed using this most general form of the CF basis, it can be greatly simplified for the ring laser geometry we consider here. Assuming further a complex but uniform refractive index $n(\tbar{\bm{r}}) \to n$, the eigenproblem can be simplified to the effectively one-dimensional case:
\begin{align}
    \tbar{\nabla}^2 \varphi_m(\tbar{x}) = -n^2 \tbar{\omega}_m^2\varphi_m(\tbar{x})
    \label{eq:helmholtz}
\end{align}
where $\tbar{x}$ denotes the angular variable along the ring cavity, and the modes satisfy periodic boundary conditions. The modes satisfy the orthogonality relationship:
\begin{align}
    \int_0^1 d\tbar{x}~\varphi^*_n(\tbar{x})\varphi_m(\tbar{x}) = \delta_{nm}
    \label{eq:orth}
\end{align}
where the integral is defined over the spatial domain of the ring cavity.

The modes obtained by solving the resulting Eq.~(\ref{eq:helmholtz}) and imposing periodic boundary conditions then take the simple form of propagating plane waves:
\begin{align}
    \varphi_m(\tbar{x}) = e^{i \tbar{k}_m \tbar{x}}e^{i(n_{\rm R}\tbar{\Omega})\tbar{x}}
\end{align}
where $\tbar{k}_m = 2m\pi, m \in \mathbb{Z}$ is the wavevector for the mode indexed by integer $m$. The complex eigenfrequencies $\tbar{\omega}_m$ can be found exactly, and are parameterized in terms of their real parts $\tbar{\nu}_m$ describing the mode frequencies and imaginary parts $\tbar{\kappa}_m$ describing losses:
\begin{align}
    \tbar{\omega}_m = \tbar{\nu}_m - i\tbar{\kappa}_m,~~~~\tbar{\nu}_m = \frac{1}{n_{\rm R}}\tbar{k}_m,~~~~\tbar{\kappa}_m = \frac{\tbar{k}_m}{n_{\rm R}} \frac{n_{\rm I}}{n_{\rm R}}. 
\end{align}
The cold cavity mode spacing $\tbar{\Delta}$ (or free spectral range (FSR)) is constant and is given by $\tbar{\Delta} = \frac{2\pi }{n_{\rm R}}$. It is then clear that the orthogonality relationship of Eq.~(\ref{eq:orth}) is simply that of the complex Fourier basis. We note that while we consider here the special case of modes of a ring cavity with only internal losses, the time-dependent theory we discuss holds for non-Hermitian modes defined by Eq.~(\ref{eq:helmholtz}) for arbitrary geometries in multiple dimensions, including random lasers~\cite{T_reci_2008}, and incorporating losses via open boundary conditions~\cite{ge_2010}. 

Having defined our spatial basis modes, we will now return to the lasing cavity case and expand the coupled slowly-varying envelopes of the electric field and polarization using the following ans\"atze:
\begin{subequations}
\begin{align}
    \tbar{E}(\tbar{x},\tbar{t}) = \sum_{m} \tbar{E}_m(\tbar{t})\varphi_m(\tbar{x}) \label{eq:ansE} \\
    \tbar{P}(\tbar{x},\tbar{t}) = \sum_{m} \tbar{P}_m(\tbar{t})\varphi_m(\tbar{x}) \label{eq:ansP}
\end{align}
\end{subequations}
The spatial complexity of the laser cavity including boundary conditions is entirely captured by the cavity modes $\{\varphi_m(x)\}$, while the nontrivial time dynamics are encoded in the expansion coefficients $\{\tbar{E}_m(t),\tbar{P}_m(t)\}$. We then substitute Eqs.~(\ref{eq:ansE}),~(\ref{eq:ansP}) into the MBEs (Eqs.~(\ref{eq:MBE})-(\ref{eq:MBD})), and make use of the orthogonality relationship to integrate out the spatial degrees of freedom over the ring cavity domain. Doing so also projects the inversion onto a set of the basis modes via:
\begin{align}
    \tbar{D}_{nm}(\tbar{t}) = \int_{0}^{1} d\tbar{x}~{\varphi^*_m}(\tbar{x})\tbar{D}(\tbar{x},\tbar{t}){\varphi_n}(\tbar{x}).
    \label{eq:ansD}
\end{align}
For the special case of a ring cavity, using the explicit form of the spatial basis modes, it is clear that $\tbar{D}_{nm}(t)$ are simply the spatial Fourier components at the wavevector \textit{difference} $\tbar{k}_n-\tbar{k}_m$ of the inversion field at a time $\tbar{t}$; however, the projection and resulting dynamical equations are more general and hold for basis modes that are not simply complex exponentials.

Leaving details of the spatial projection for Appendix~\ref{app:cftd}, we present the final equations of motion for the variables $\{\tbar{E}_m,\tbar{P}_m,\tbar{D}_{nm}\}$ below:
\begin{subequations}
\label{eq:CFTD_set}
\begin{align}
    \dot{\tbar{E}}_m &= \frac{i}{2\tbar{\Omega}}\left(\tbar{\Omega}^2 - \tbar{\nu}_m^2 + \tbar{\kappa}_m^2 \right)\tbar{E}_m - \tbar{\kappa}_m\tbar{E}_m +\frac{i\tbar{\Omega}}{2n_{\rm R}^2}\tbar{P}_m \label{eq:ET}  \\
    \dot{\tbar{P}}_m &= -\tbar{\gamma}_{\perp} \tbar{P}_m - i\tbar{\gamma}_{\perp} \sum_m \tbar{E}_n\tbar{D}_{mn}  \label{eq:PT} \\
    \dot{\tbar{D}}_{nm} &= -\tbar{\gamma}_{\parallel}(\tbar{D}_{nm}-\tbar{D}^0_{nm}) + \frac{i\tbar{\gamma}_{\parallel}}{2}\!\sum_{rs}\! \tbar{\mathcal{A}}_{nmrs}\!\left[\tbar{E}_{r}\tbar{P}_{s}^* - \tbar{E}_{s}^*\tbar{P}_{r}\right] \label{eq:DT}
\end{align}
\end{subequations}
where we have introduced the dimensionless mode overlap tensor $\tbar{\mathcal{A}}_{nmrs}$:
\begin{align}
    \tbar{\mathcal{A}}_{nmrs} &= \int_0^1 d\tbar{x}~\varphi_n(\tbar{x})\varphi^*_m(\tbar{x})\varphi_r(\tbar{x})\varphi^*_s(\tbar{x}) \nonumber \\
    &=  \int_0^1 d\tbar{x}~e^{i(\tbar{k}_n+\tbar{k}_r-\tbar{k}_m-\tbar{k}_s)\tbar{x}} = \delta(n+r-m-s)
    \label{eq:Adef}
\end{align}
The second line above specializes the tensor to the case of multimode ring lasers. Finally, we note that $\tbar{D}^0_{nm}$ defines the projected matrix elements for the incoherent pump, and is defined analogously to Eq.~(\ref{eq:ansD}), with $\tbar{D}(x,t) \to \tbar{D}^0$. This definition can equivalently be used for spatially inhomogeneous pump profiles.

Eqs.~(\ref{eq:ET})-(\ref{eq:DT}) thus define the CFTD description of multimode dynamics for ring lasers: a time-domain description derived from the underlying spatiotemporal MBEs via a specialized spatial projection, most generally using the CF basis. This leads to a close connection between CFTD and SALT, which we expand on below.


\subsection{Dynamics across regimes of stationary and non-stationary inversion}
\label{subsec:cftdDyn}

The assumption of stationary inversion, $\dot{\mathcal{D}}(t) = 0$, forms the basis of successful steady-state spectral descriptions of lasing, such as SALT. However, this assumption places strong constraints on lasing phenomena described by the MBEs. To see this, note that if the inversion is stationary Eqs.~(\ref{eq:MBE})-(\ref{eq:MBP}) for the electric and polarization fields form a set with no nonlinear mixing of \textit{time-dependent} terms. While the inversion does depend nonlinearly on $\mathcal{E}$ and $\mathcal{P}$ via Eq.~(\ref{eq:MBD}), it is assumed to be time-independent and hence does not lead to any frequency mixing in Eqs.~(\ref{eq:MBE})-(\ref{eq:MBP}). The same observation holds for the subsequently-derived CFTD equations. Stationary inversion thus precludes the generation of new frequencies via intermodulation products of the lasing field and the gain medium population dynamics (which are strongly suppressed). New frequencies can arise in this regime, but via standard multimode lasing: with increasing pump power, a new spatial mode $\varphi_m$ can lase if a corresponding pump threshold is crossed. These pump thresholds provide a natural truncation scheme within SALT, determining when a specific mode begins to lase and hence must be included in the spectral description.


In contrast, CFTD is not bound by the stationary inversion approximation, and is able to simulate dynamics across regions of both stationary \textit{and} non-stationary inversion. For aforementioned regimes where SALT is valid, the CFTD ansatz can capture lasing modes $\varphi_m$ with nonzero coefficients $\tbar{E}_m(t)$ containing only a single frequency in the long-time limit, defining the lasing frequency for that mode $m$. However, by allowing $\tbar{E}_m(t)$ to have more a general time dependence, CFTD can also capture \textit{transient} dynamics where the inversion evolves with time, to the final steady-state lasing regime where SALT operates; we discuss this dynamics in Sec.~\ref{sec:singleModeLasing}. We note that correspondence between CFTD and SALT in regimes where the stationary inversion approximation is briefly violated were also investigated in Ref.~\cite{malik_spectral_2015}.

Most importantly, in this paper we analyze regimes where the dynamics of the inversion give rise to lasing phenomena that are \textit{qualitatively} different from multimode lasing: instabilities that lead to simultaneous generation of multiple coherent frequencies from a single lasing frequency background. The coefficients $\tbar{E}_m(t)$ associated with distinct spatial modes now have much more general, coupled dynamics: several coefficients can become nonzero simultaneously past a single threshold, instead of sequentially with multiple thresholds. Each $\tbar{E}_m(t)$ can even possess multiple distinct frequency components, a feature increasingly prevalent in more complex laser geometries~\cite{malik_nonlinear_2017}. This more general class of dynamics is the main focus of this paper, and the subject of Secs.~\ref{sec:rngh} and~\ref{sec:chaos}. 



\subsection{Numerical verification}

We investigate the accuracy and efficiency of the CFTD model by comparing against standard numerical approaches to simulating Eqs.~(\ref{eq:SVE})-(\ref{eq:SVD}). The most natural comparison of the CFTD method, which is geometry-agnostic, is against a completely general spatiotemporal FDTD scheme. For specific comparisons, however, we also employ the split-step Runge-Kutta method (SSRK)~\cite{fornberg_fast_1999, hult_fourth-order_2007}. This scheme is tailored to ring lasers, and hence can be expected to perform optimally for the present case, but lacks generalizability to other geometries. We begin with comparisons in both simple single-mode lasing regimes but with nontrivial initial spatial field distributions (i.e. spanning multiple cavity modes) in Sec.~\ref{sec:singleModeLasing}. When this single-mode lasing state becomes unstable, complex multimode lasing dynamics can ensue, including the formation of frequency combs which we analyze in Sec.~\ref{sec:rngh} or even the emergence of chaotic dynamics, discussed in Sec.~\ref{sec:chaos}.

\section{Single-mode lasing dynamics}
\label{sec:singleModeLasing}

We will begin our analysis of lasing dynamics with the simplest operating regime: single-mode lasing. Within our spectral approach, the single-mode lasing regime is characterized by restricting the expansion coefficients in Eqs.~(\ref{eq:ansE}),~(\ref{eq:ansP}) to the single-mode forms:
\begin{align}
    \tbar{E}_{m}(t) &= \delta_{lm}\tbar{E}_{l}(t) \nonumber \\
    \tbar{P}_{m}(t) &= \delta_{lm}\tbar{P}_{l}(t)
\end{align}
so that the only mode with nonzero expansion coefficients is indexed by $l$, and is thus the solitary mode that lases. By virtue of how the inversion matrix elements are constructed - as projections onto the modes constituting the expansion of Eqs.~(\ref{eq:ansE}),~(\ref{eq:ansP}) - the resulting inversion matrix elements, Eqs.~(\ref{eq:ansD}), similarly reduce to:
\begin{align}
    \tbar{D}_{nm}(t) = \delta_{ln}\delta_{lm}\tbar{D}_{ll}(t)
\end{align}

As mentioned earlier, we consider a lasing cavity that experiences a spatially uniform pump gain and uniform loss profile. This is not a restriction of CFTD, but is a simplification we make for convenience of later comparisons. Under this assumption, it is easily found that the mode with lowest threshold pump power is one that is spectrally closest to the atomic transition frequency $\tbar{\Omega}$ (for details see Appendix~\ref{app:sml}),; we index this mode by $l=0$, so by definition $|\tbar{\nu}_{0}-\tbar{\Omega}| \ll |\tbar{\nu}_{l}-\tbar{\Omega}|~\forall~l~\neq 0$. We emphasize, however, that this simple result holds provided the gain and loss distribution in the laser cavity is uniform. To simplify the analysis further, we consider a cavity such that $\tbar{\nu}_{0} \simeq \tbar{\Omega}$, in which case the lasing occurs at the frequency $\tbar{\Omega}$. This frequency is explicitly extracted in the CFTD ans\"atze, Eq.~(\ref{eq:ansE}),~(\ref{eq:ansP}), so that the lasing mode in this frame is at zero frequency. As a result, the steady-state fields $\tbar{E}_{0} = \ESS, \tbar{P}_{0} = \PSS, \tbar{D}_{00} = \DSS$ in the single-mode lasing regime are all stationary, and are given by:
\begin{align}
    \DSS = \DthT = \frac{2n_{\rm R}^2\tbar{\kappa}}{\tbar{\Omega}},~|\ESS|^2 = \frac{\tbar{D}^0_{00}}{\DthT} - 1,~\PSS = -i\ESS\DSS 
\end{align}
Details of the derivation of the above expression can be found in Appendix~\ref{app:sml}.


\begin{figure}[t]
    \centering
    \includegraphics[scale=1.0]{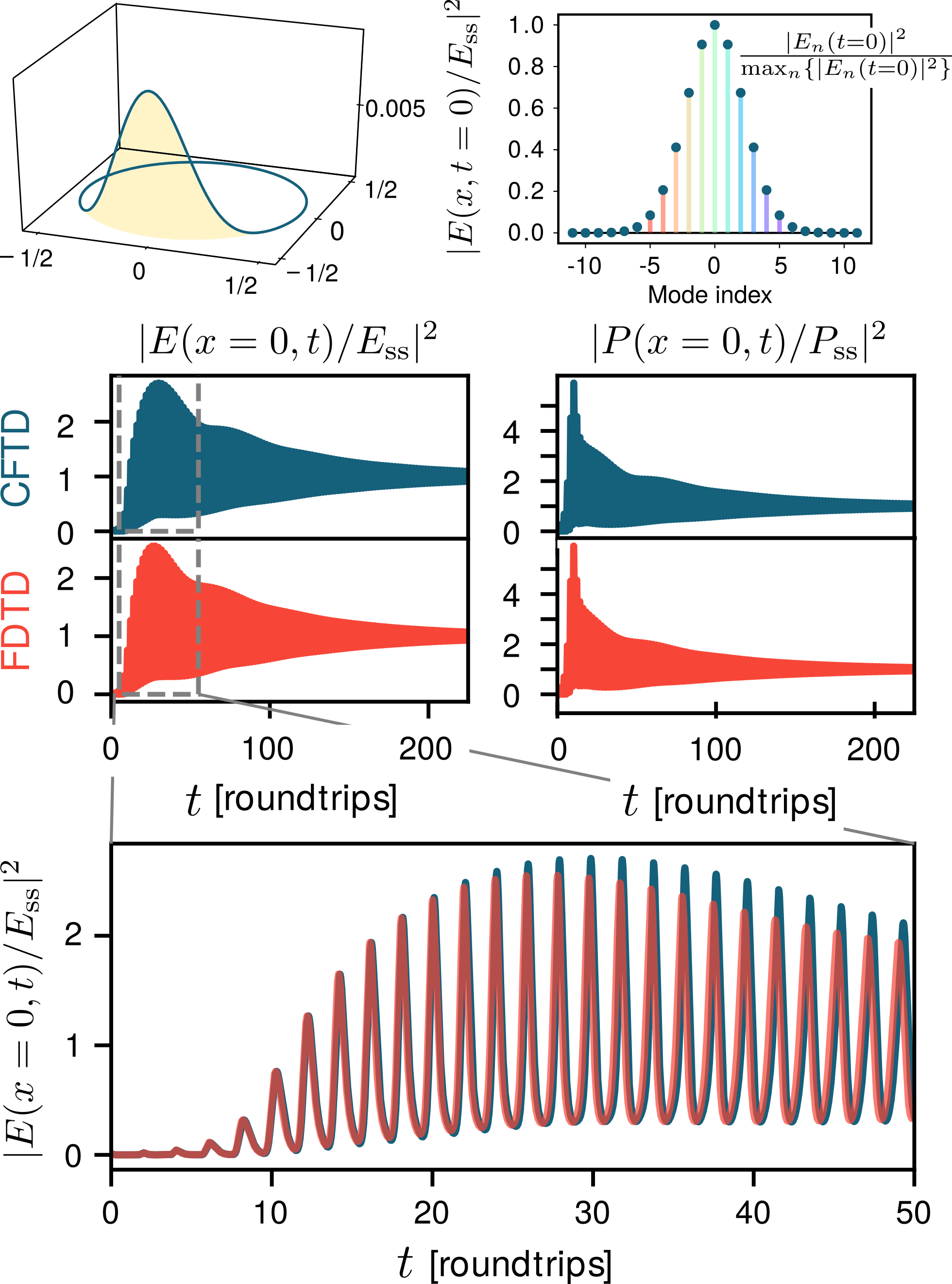}
    \caption[]{Multimode dynamics below the instability threshold (the pump power normalized by the single mode lasing threshold is $p = 5$). Top left panel shows the initial Gaussian electric field profile, $\tbar{E}(x,t=0) = 0.1 e^{-100x^2}$, scaled by $\ESS$. Projection of this initial profile on the CFTD spatial basis is shown in the top right plot. Center panel: comparison of dynamics of the electric field intensity $|E(x=0,t)|^2$ and polarization $|P(x=0,t)|^2$ using CFTD (blue) and FDTD (red). Lower panel shows $|\tbar{E}(x=0,t)|^2$ scaled by $\ESS$ for the first few roundtrips to highlight the agreement. Loss parameters are $\tbar{\gamma}_{\perp}=5.0,\tbar{\gamma}_{\parallel}=0.5,\tbar{\kappa}=0.1$, while $n_{\rm R}=1.96$. The number of modes included is $N = 11$, the minimum for reproducing an accurate Gaussian profile as the initial condition.} 
    \label{fig:BelowThreshG5}
\end{figure}


In this single-mode regime, it would appear that a spectral approach retaining only the mode that eventually lases would suffice, namely restricting Eqs.~(\ref{eq:ansE})-(\ref{eq:ansD}) to $m=n=l=0$. However, our more general ans\"atze allows us to quantitatively capture dynamics that require modes beyond the lasing mode, for example pulsed initial conditions or spatially non-uniform pumping schemes that excite modes other than the lasing mode, even if such modes decay away in the long-time limit.

To simulate this nontrivial regime, we explore the multimode dynamics of a ring laser pumped above the single-mode lasing threshold, with an initial intensity distribution within the laser cavity that is described by a Gaussian profile. We consider pumping the laser at a power five times above the single-mode lasing threshold, $p=\frac{\tbar{D}^0_{00}}{\DthT} = 5$, with decay parameters $\tbar{\gamma}_{\perp}=5.0,\tbar{\gamma}_{\parallel}=0.5,\tbar{\kappa}=0.1$, cold cavity refractive index set to $n_{\rm R}=1.96$, and then simulate the dynamics of the electric and polarization fields at a fixed position $\tbar{x} = 0$ of the ring cavity as a function of time, using both FDTD simulations of Eqs.~(\ref{eq:SVE})-(\ref{eq:SVD}), and integration of the CFTD Eqs.~(\ref{eq:ET})-(\ref{eq:DT}). The CFTD approach here takes into consideration modes $m \in {-5,\ldots,5}$ for a total of $N=11$ modes. A time-step $\Delta t = 2.45 \times 10^{-4}$ in units of roundtrips is used for FDTD simulations, while CFTD employs an ODE solver with an adaptive time-step.

The results are plotted in Fig.~\ref{fig:BelowThreshG5}, in red for the FDTD simulations and blue for the CFTD simulations.  We find excellent quantitative agreement between the two approaches; the lower panel zooms in on a length of time equal to 50 roundtrips, plotting both CFTD and FDTD results to highlight the agreement. Under stationary inversion, only a single-mode solution exists (see Appendix~\ref{app:sml}). However, the nontrivial time evolution indicates multimode dynamics in a transient period of $\sim$ 200 roundtrips after which a single mode lases in the long-time limit. The CFTD dynamics are substantially more efficient to simulate than the FDTD, requiring simulation times that are about two orders-of-magnitude shorter than the FDTD for the same timestep in regimes captured by several cavity modes (a more detailed comparison of CFTD simulation times versus the number of retained modes in more complex dynamical regimes is shown in Sec.~\ref{sec:speedup}).


\begin{figure}[t]
    \centering
    \includegraphics[scale=1.0]{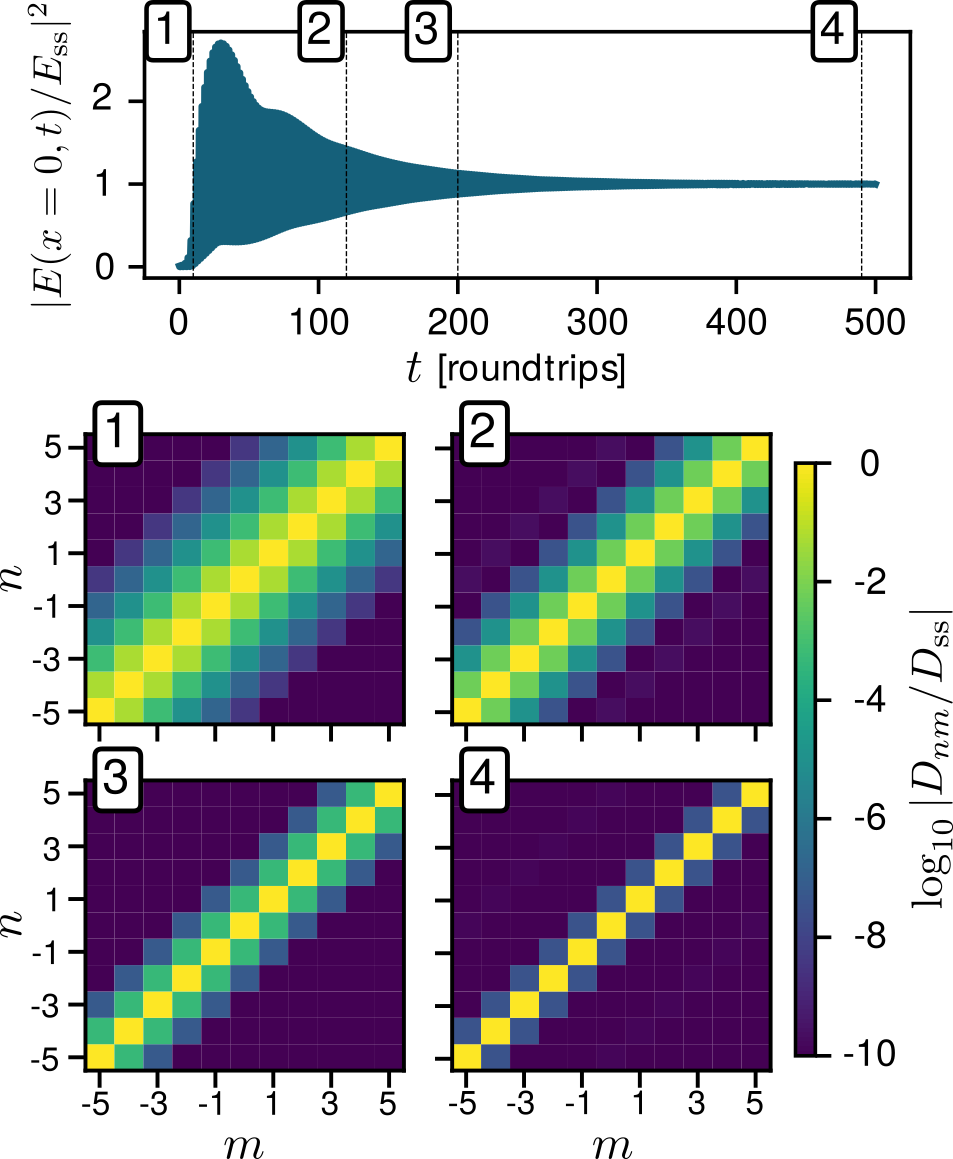}
    \caption[]{Top panel: Evolution of the electric field intensity in the single-mode lasing regime for 500 roundtrips. Bottom panel: Dynamics of inversion matrix elements $\tbar{D}_{nm}(t)$ in CFTD, evaluated at labelled time points, below the single-mode instability threshold ($p = 5$) for an initial Gaussian electric field profile corresponding to Fig.~\ref{fig:BelowThreshG5}. Parameters are the same as that figure. Off-diagonal elements of the matrix are suppressed as the solution reaches the single-mode steady state.  }
    \label{fig:Dmatrix}
\end{figure}


A unique aspect of the CFTD approach is the use of a projected set of inversion matrix elements $\tbar{D}_{nm}(t)$, Eq.~(\ref{eq:ansD}), to capture the dynamics of the inversion field. Formally, this projection cannot be straightforwardly inverted to extract the spatio-temporal inversion field $D(x,t)$. For the case of a ring laser, these matrix elements have a simple physical interpretation. For a spatially-uniform inversion field, $D(x,t) = D(t)$, all off-diagonal inversion matrix elements vanish, as is seen from Eq.~(\ref{eq:ansD}) and the orthogonality relationship, Eq.~(\ref{eq:orth}). This case represents the single-mode lasing regime for ring lasers. However, when multiple spatial modes with distinct wavevectors are active, a spatial inversion grating is established in the laser cavity~\cite{mansuripur_single-mode_2016}: $\tbar{D}_{nm}(t)$ then precisely represent the time-evolution of spatial Fourier components of the inversion field at the \textit{difference} of wavevectors $k_n-k_m$. For more general cavities where the projecting modes $\{\varphi_m(x)\}$ are not simply Fourier basis elements, $\tbar{D}_{nm}(t)$ for $n \neq m$ still represents the amplitude of this inversion grating.

The evolution of the inversion matrix elements can be seen in Fig.~\ref{fig:Dmatrix}, where $\tbar{D}_{nm}(t)$ are plotted in the 2-D plots at indicated time values 1 through 4 scaled by the single-mode lasing steady-state inversion, $\DSS$. For the first $\approx 200$ roundtrips the multimode initial pulse is travelling through and in fact building up in the ring laser, and this dynamics leads to the generation of off-diagonal inversion matrix elements due to the inversion grating. However, with increasing time, additional modes decay away and the laser returns to a single-mode lasing state; concurrently, the off-diagonal inversion matrix elements are suppressed. 

We thus see that the CFTD can accurately capture spatiotemporally-complex initial conditions in the regime of single-mode lasing. With increasing pump power, it is natural to ask whether the single-mode lasing regime analyzed in this section breaks down, and if other cavity modes can be coherently excited in the long-time limit, unlike the dynamics observed in Fig.~\ref{fig:BelowThreshG5}. This is the subject of the next section.


\section{Coherent Multimode lasing dynamics: RNGH instability}
\label{sec:rngh}

It can be shown that uniformly incoherently-pumped ring lasers experience gain clamping above the single-mode lasing threshold. The inversion field attains a stationary, spatially-homogeneous value, and prevents additional modes from crossing the lasing threshold while the first mode is still lasing. However, this does not mean that multi-mode dynamics cannot ensue: other types of instabilities do exist, where the single lasing mode itself becomes unstable, leading to the generation of new frequencies mediated by the gain medium. Such a regime is marked by the inversion field becoming time-dependent, leading to so-called \textit{population pulsations}~\cite{pop_pulsations}.

\begin{figure}[t]
    \centering
    \includegraphics[scale=1.0]{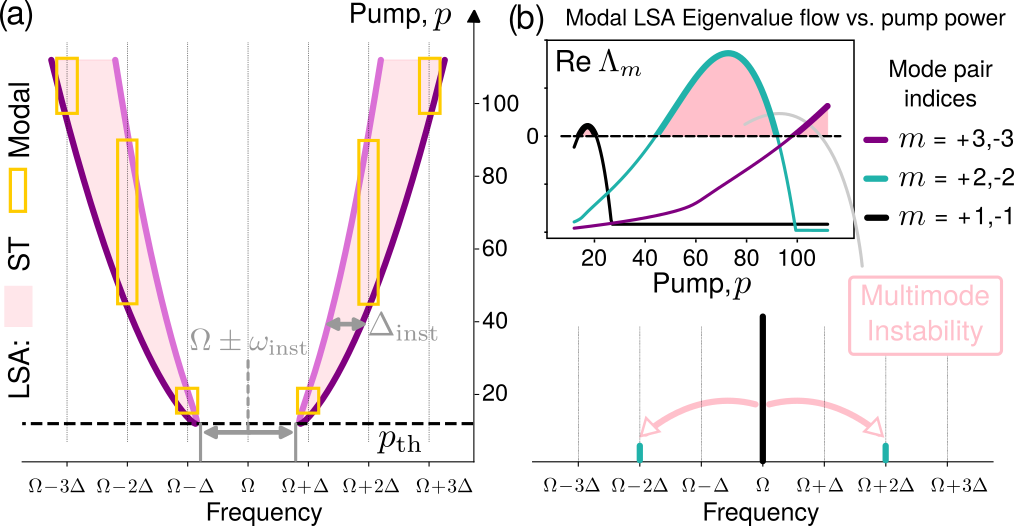}
    \caption[]{(a) ST LSA results based on the original formulation of Risken \textit{et al.}~\cite{risken_selfpulsing_1968}, and corresponding modal LSA results. Beyond a specific pump threshold $p_{\rm th}$ [See Eq.~(\ref{eq:rnghpth})], the single-mode lasing solution becomes unstable, at a frequency given by $\tbar{\omega}_{\rm inst}$. Dashed lines indicate the results of the modal LSA based on linearization of the CFTD equations. (b) Real part of the eigenvalues $\Lambda_m$ of the dynamical matrix in Eq.~(\ref{eq:jacobian}) as a function of pump power. A positive real part indicates an instability. Specific ring laser parameters considered here are $\tbar{\gamma}_{\perp}=1.0,\tbar{\gamma}_{\parallel}=0.5,\tbar{\kappa}=0.1$, and $n_{\rm R}=1.96$.}
    \label{fig:RNGHResults}
\end{figure}

The LH and RNGH instabilities~\cite{risken_selfpulsing_1968, graham_quantum_1968} represent prototypical examples of such a strongly-pumped, nonlinear multimode lasing regime. The salient features of these instabilities were originally predicted via a spatiotemporal linear stability analysis (ST LSA) of the MBEs~\cite{risken_selfpulsing_1968} and then verified numerically; these features are summarized in Fig.~\ref{fig:RNGHResults} and discussed below. Above a critical (normalized) pump power that we refer to as the \textit{instability threshold} $p_{\rm th}$ (the so-called second threshold in early literature),
\begin{align}
    p_{\rm th} = 5 + 3\left(\frac{\tbar{\gamma}_{\parallel}}{\tbar{\gamma}_{\perp}} \right) + 2 \sqrt{4 + 6\left(\frac{\tbar{\gamma}_{\parallel}}{\tbar{\gamma}_{\perp}}\right) + 2\left(\frac{\tbar{\gamma}_{\parallel}}{\tbar{\gamma}_{\perp}}\right)^{\ 2} }, 
    \label{eq:rnghpth}
\end{align}
the single-mode lasing solution becomes unstable (see red dashed red line in Fig.~\ref{fig:RNGHResults}), giving rise to symmetric sidebands at the frequencies $\Omega\pm\tbar{\omega}_{\rm inst}$. At higher pump powers, a continuous range of frequencies as marked by the shaded region are predicted to be unstable; the width of this region in frequency is given by $\tbar{\Delta}_{\rm inst}$. 

Here, we note the distinction that is often made in the literature, according to the ratio of the gain linewidth $\gamma_{\perp}$ to the FSR $\Delta$. In particular, for $\gamma_{\perp}/\Delta < 1$, a single cavity mode - the lasing mode - falls under the gain curve. The resulting instability is referred to as the single-mode or LH instability. In the opposite regime where $\gamma_{\perp}/\Delta > 1$, several cavity modes fall under the gain curve; consequently, the emerging instability is labelled the multi-mode or RNGH instability. While dynamics in both regimes can be distinct (as we will see), both are described by the same ST LSA, with different parameters.

\sakc{In specific parameter regimes, these ST LSA results can be cast in simplified analytic forms. In particular, in the RNGH case and further considering $\gamma_{\perp} \gg \Delta,\gamma_{\parallel},\kappa$,} the minimum unstable frequency and the range of unstable frequencies close to the \sakc{instability threshold $p_{\rm th}$} take the forms~\cite{risken_selfpulsing_1968},
\begin{subequations}
\begin{align}
    \pm\tbar{\omega}_{\rm inst}  &= \pm \tbar{\gamma}_{\perp}\left(2\sqrt{3}-\frac{2}{\sqrt{3}}\frac{\tbar{\kappa}}{\tbar{\gamma}_{\perp}}\right)\sqrt{\frac{\tbar{\gamma}_{\parallel}}{\tbar{\gamma}_{\perp}}}  \label{eq:wmin} \\
    \tbar{\Delta}_{\rm inst} &\simeq \tbar{\gamma}_{\perp}\left(\frac{\sqrt{6}}{2}+\frac{7}{\sqrt{6}}\frac{\tbar{\kappa}}{\tbar{\gamma}_{\perp}}\right)\sqrt{\frac{\tbar{\gamma}_{\parallel}}{\tbar{\gamma}_{\perp}}}\sqrt{\frac{p}{p_{\rm th}}-1} \label{eq:deltainst}
\end{align}
\end{subequations}


Note that both $\omega_{\rm inst}$ and $\Delta_{\rm inst}$ generally grow with $\gamma_{\perp}$. While the magnitude of this linewidth in comparison to the FSR ($\gamma_{\perp}/\Delta$) is heuristically understood to increase the participation of cavity modes in lasing, here we see its role in increasing the participation of modes in multimode instabilities. As we will see via numerical simulations, this increase in $\Delta_{\rm inst}$ via $\gamma_{\perp}/\gamma_{\parallel}$ will lead to a growing complexity of multimode dynamics.

Having described the general multimode instability, we now discuss how the CFTD approach can also predict this instability using an efficient modal linear stability analysis (\textit{modal} LSA).





\subsection{Modal LSA: instability via non-stationary inversion}

Within the CFTD approach, instability of the single-mode lasing solution arises naturally via the growth of unstable sidebands for $m \neq 0$. We analyze the dynamics of these sidebands in a linearized approximation:
\begin{subequations}
\begin{align}
    \tbar{E}_m &= \tbar{E}_{\rm ss}\delta_{0m} +  \delta{E}_m + O(\delta \tbar{E}_m^2) \\
    \tbar{P}_m &= \tbar{P}_{\rm ss}\delta_{0m} + \delta{P}_m + O(\delta \tbar{P}_m^2)  \\
    \tbar{D}_{nm} &= \tbar{D}_{\rm ss}\delta_{nm} + \delta \tbar{D}_{nm} + O(\delta \tbar{D}_{nm}^2)
    \label{eq:linAnsatze}
\end{align}
\end{subequations}
and retaining only terms to linear order in $\{\delta\tbar{E}_m,\delta\tbar{P}_m,\delta\tbar{D}_{nm}\}$.


The multimode LSA then proceeds by first obtaining linearized dynamical equations for the sideband fluctuations and inversion matrix elements. Leaving details of the derivation for Appendix~\ref{app:stability}, the resulting equations can be conveniently written in matrix form as:
\begin{align}
    \frac{d}{dt}\!\!
    \begin{pmatrix}
    \delta\vec{v}_m \\
    \delta\vec{v}_{-m}
    \end{pmatrix}
    \equiv \mathcal{M}_m\!\!     \begin{pmatrix}
    \delta\vec{v}_m \\
    \delta\vec{v}_{-m}
    \end{pmatrix}= 
    \begin{pmatrix}
    \mathbf{J}_m & \mathbf{C}_{m,-m} \\
    \mathbf{C}_{m,-m} & \mathbf{J}_{-m}
    \end{pmatrix}
    \!\!
    \begin{pmatrix}
    \delta\vec{v}_m \\
    \delta\vec{v}_{-m}
    \end{pmatrix}
    \label{eq:jacobian}
\end{align}
where the component vectors $\delta\vec{v}_m$ include fluctuation variables for mode $m$:
\begin{align}
    \vec{v}_m = (\delta\tbar{E}_{m},\delta\tbar{E}_{m}^*,\delta\tbar{P}_{m},\delta\tbar{P}_{m}^*,\delta\tbar{D}_{0m},\delta\tbar{D}_{0m}^* )^T
\end{align}
Immediately, we note that a closed set of linearized equations can only be obtained when considering the coupled dynamics of \textit{pairs} of modes $\{m,-m\}$. Here $\mathbf{J}_{m}$ is a 6-by-6 dynamical matrix of the fluctuations of mode $m$ alone, including those of the inversion matrix elements. Importantly, the dynamics due to $\mathbf{J}_m$ alone lead to a decay of the $m$th mode fluctuations in the long-time limit, so that the single-mode lasing solution ($m=0$) remains stable. 

However, the matrix $\mathbf{C}_{m,-m}$ couples fluctuations of different modes, in particular the symmetric mode pair $\{m,-m\}$. As $\mathbf{J}_m$ does not lead to instability, any multimode instabilities arise due to this coupling matrix. Crucially, as we show in Appendix~\ref{app:stability}, the coupling is mediated by nonzero \textit{off-diagonal} inversion matrix elements $D_{0m}$, namely $\pm m \neq 0$. For a ring laser with \textit{stationary} inversion, the inversion is also spatially homogeneous, and from Eq.~(\ref{eq:ansD}) will thus only lead to nonzero \textit{diagonal} inversion matrix elements $D_{mm}$. As a result, the requirement of $D_{0m} \neq 0$ for nonzero $m$ in a ring laser necessitates non-stationary inversion, accounted for within the framework of CFTD.
 
We now present results of the CFTD stability analysis enabled by Eqs.~(\ref{eq:jacobian}), which evaluates the stability of each mode pair $\{m,-m\}$ via the eigenvalues of the dynamical matrix $\mathcal{M}_m$. As an example, we plot the largest real part of the eigenvalue spectrum, ${\rm Re}~\Lambda_m$, for a selection of mode pairs in Fig.~\ref{fig:RNGHResults}, as a function of pump power. We see that with increasing pump power, the eigenvalue for a specific mode pair can cross the instability threshold (its real part becomes positive). These shaded regions correspond to mode pairs experiencing gain and growing around the single-mode lasing solution, as depicted in the lower panel. However, note that the eigenvalue flow can be highly monotonic, so that a specific mode pair can become sub-threshold again with increasing pump power. At any pump power, the predicted unstable mode pairs are marked with solid lines in Fig.~\ref{fig:RNGHResults}(a), where they coincide perfectly with the RNGH instability region for the parameters considered.

Importantly, pump powers also exist where the CFTD modal LSA predicts no unstable mode pairs. For such pump powers, the RNGH stability analysis predicts an unstable frequency, but this frequency does not coincide with that of a cold cavity mode (dotted vertical lines). The CFTD analysis indicates that for such pump powers, there will be no multimode instability; we will verify these predictions numerically in the following section.

The modal LSA enabled by CFTD therefore agrees well with the original RNGH analysis for the considered parameters, but is based on a more general expansion that could be applied to studying multimode instabilities in alternative laser geometries. 

\begin{figure*}[t]
    \centering
    \includegraphics[scale=0.78]{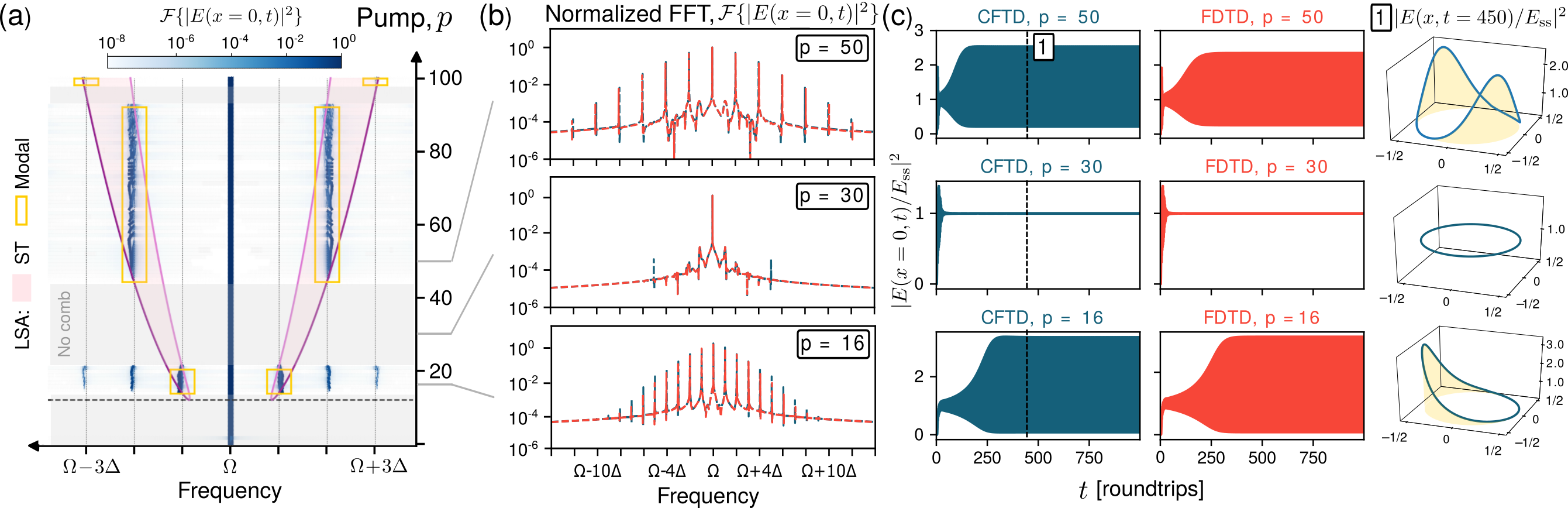}
    \caption[]{Emergence of frequency combs via the LH instability in ring lasers for loss parameters $\tbar{\gamma}_{\perp}=1.0,\tbar{\gamma}_{\parallel}=0.5,\tbar{\kappa}=0.1$, and $n_{\rm R}=1.96$. (a) Frequency spectrum of the electric field as a function of pump power obtained using CFTD. Shaded regions mark the continuous frequencies predicted to be unstable by the ST LSA, while yellow boxes mark the discrete mode pairs predicted to be unstable by the modal LSA. Cold cavity mode frequencies are shown in dashed black, while the horizontal dashed line is the instability pump threshold $p_{\rm th}$. (b) Normalized electric field spectra at specific pump powers, and (c) their corresponding time traces $|E(x=0,t)|^2$ scaled by $E_{\rm ss}$, computed using both CFTD (blue) and FDTD (red) for a direct comparison. Right panel: Spatial distribution of the steady-state pulsing electric field in the ring cavity at a fixed time, $|E(x,t=450)/E_{\rm ss}|^2$ (marked by (1) in (c)), computed using CFTD across the three pump powers simulated in the previous panels.}
    \label{fig:mainfig}
\end{figure*}


\subsection{Good cavity regime and slow polarization relaxation $\gamma_{\perp}/\gamma_{\parallel} \lesssim 2.0$}

We are now in a position to simulate the dynamics of ring lasers in parameter and pump regimes where both the approximate ST stability analysis and its modal counterpart enabled by CFTD predict multimode instabilities. We will do so by integrating the exact CFTD equations, Eqs.~(\ref{eq:CFTD_set}), and compare the results against integration of the MBEs under the slowly-varying envelope approximation, Eqs.~(\ref{eq:SVE})-(\ref{eq:SVD}). To begin, we focus on the ``good cavity'' regime where $\tbar{\kappa} < \tbar{\gamma}_{\perp}+\tbar{\gamma}_{\parallel}$, considering $\tbar{\gamma}_{\perp}=1.0,\tbar{\gamma}_{\parallel}=0.5,\tbar{\kappa}=0.1$ and $n_{\rm R} = 1.96$; dynamics in the ``bad cavity'' limit are analyzed in Sec.~\ref{sec:chaos}. \sakc{Note that $\gamma_{\perp}<\Delta = \frac{2\pi}{n_{\rm R}}$, so the emergent instability would typically be characterized as of the LH type.} Finally, as the carrier and polarization decay rates are similar (hence, their ratio is relatively small), this regime may be representative of so-called class A lasers, depending on the choice of $\kappa$. This class of lasers includes dye lasers, HeNe lasers and quantum cascade lasers.




We simulate the CFTD equations to obtain $\{E_m(t),P_m(t),D_{nm}(t)\}$ for $m\in -5,\ldots,5$ (i.e. $N=11$ spatial modes). This allows us to reconstruct the electric field using Eq.~(\ref{eq:ansE}), namely $E(x,t) = \sum_m E_m(t)\varphi_m(x)$, as a function of normalized pump powers $p$ over a large range past the single mode lasing threshold, evolving for $t=1000$ roundtrips at each pump power. This detailed study is greatly aided by the numerical efficiency of the CFTD equations, which provide a significant simulation speedup in comparison to spatiotemporal schemes like FDTD (see Sec.~\ref{sec:speedup}). We first extract the resulting normalized frequency spectrum of the electric field at $x=0$, $\mathcal{F}\{E(x=0,t)/E_{\rm ss}|^2\}$, which is shown in Fig.~\ref{fig:mainfig}(a), with amplitudes depicted according to the listed color scale. 

Until the RNGH threshold $p_{\rm th}$ is reached, a single-mode lasing regime is clearly observed. Past this threshold, the predicted instability of single mode lasing dynamics is observed, giving way to a stable waveform exhibiting a frequency comb in its electric field frequency spectrum. Importantly, the dominant unstable sidebands occur at $\Omega\pm\Delta$, consistent with the prediction of both LSA methods. This instability propagates to a broad frequency comb with a spacing of one FSR. A more detailed look at the observed comb is presented in Fig.~\ref{fig:mainfig}(b) for $p=16$, which also shows the comparison between CFTD and FDTD, finding excellent agreement. Fig.~\ref{fig:mainfig}(c) shows the CFTD and FDTD simulations in the time domain, demonstrating the ability of CFTD to capture not only the steady-state frequency comb but also the transient approach to this state. 

Finally, the right panel of Fig.~\ref{fig:mainfig}~(c) shows the spatial electric field profile at $\tbar{t}=450$ (roundtrips), reconstructed from the CFTD solution. In this particular regime, each $E_m(t)$ associated with the spatial mode $\varphi_m$ evolves at a single dominant frequency; hence the multiple frequency components observed in Fig.~\ref{fig:mainfig}~(b) indicate a waveform that consists of several spatial modes of the cavity being coherently excited. This yields a localized propagating pulse in the laser cavity, sometimes referred to as a Turing roll in the spatiotemporal pattern formation literature~\cite{godey_stability_2014}. We verify these CFTD results against the FDTD and SSRK (not shown), finding excellent agreement once more.

With increasing pump power, the frequency comb suddenly collapses back to stable single mode lasing operation. Here, even though the ST LSA predicts a continuous instability band, no specific mode falls into this band. The modal linear stability approach, on the other hand, predicts discrete pairs of unstable modes, and here finds no such pairs to be unstable, consistent with the simulated dynamics. At a representative pump power $p=30$, the observed frequency spectrum in Fig.~\ref{fig:mainfig}(b) using both CFTD and FDTD shows no comb formation, and the time traces in Fig.~\ref{fig:mainfig}(c) exhibit no oscillations, instead settling into the single mode steady state solution $\ESS$, with a uniform spatial distribution as viewed in the right panel. 

With further increase in pump power, the frequency comb emerges again, now with a 2-FSR spacing. Here, the unstable sidebands at $\Omega\pm2\Delta$ fall in the instability region, and are also predicted to be unstable by the modal LSA. Again at the representative value of $p=50$, CFTD and FDTD show excellent agreement in capturing the extent of the frequency spectrum of the comb in Fig.~\ref{fig:mainfig}(b), oscillations in the time domain in Fig.~\ref{fig:mainfig}(c), and the spatial profile in the rightmost panel, which now displays two peaks and half the propagation period, consistent with the higher spacing of peaks in the frequency spectrum.

In summary, in this regime of slower polarization relaxation, we find excellent quantitative agreement between FDTD simulations of the MBE and our CFTD approach, as well as between the ST LSA and its modal counterpart derived from the CFTD equations. We also note that in this regime, both methods agree with the SSRK method specific to ring lasers (not shown).



\subsection{Good cavity regime and slow polarization relaxation $\gamma_{\perp}/\gamma_{\parallel} \gtrsim 20.0$}
\label{secsec:fastpolarization}

The ratio ${\tbar{\gamma}_{\perp}}/{\tbar{\gamma}_{\parallel}}$ plays an important role in the features of the ST LSA as explained by Eq.~(\ref{eq:deltainst}). Thus far we have explored regimes for which ${\tbar{\gamma}_{\perp}}/{\tbar{\gamma}_{\parallel}}\leq 2$, and have found the CFTD to provide excellent quantitative agreement with exact integration using FDTD, while providing the speedup advantages afforded by a temporal scheme as opposed to a spatiotemporal one. As the ratio  ${\tbar{\gamma}_{\perp}}/{\tbar{\gamma}_{\parallel}}$ increases, both the threshold frequency at which instability occurs and the width of the instability band increase, leading to more involved comb formation dynamics at high pump powers. In these more complex regimes, we find that the CFTD still provides very good qualitative agreement with more numerically-expensive methods. For a concrete demonstration, we consider the ratio ${\tbar{\gamma}_{\perp}}/{\tbar{\gamma}_{\parallel}}=20$, with decay parameters $\tbar{\gamma}_{\perp}=10.0,\tbar{\gamma}_{\parallel}=0.5,\tbar{\kappa}=0.1$ (for supplementary simulations with different parameters, see Appendix~\ref{app:simTime}). \sakc{Now, $\gamma_{\perp} >\Delta$, so the emergent instability would generally be characterized as the RNGH instability.} We note also that this case where the polarization decay rate is significantly larger than the population decay rate typically describes lasers in class B, depending on the value of $\kappa$~\cite{de_valcarcel_riskennummedalgrahamhaken_1999}. This class includes lasers of high technological value such as semiconductor lasers and solid state lasers.





\begin{figure}[t]
    \hspace{-0.5em}
    \includegraphics[scale=0.89]{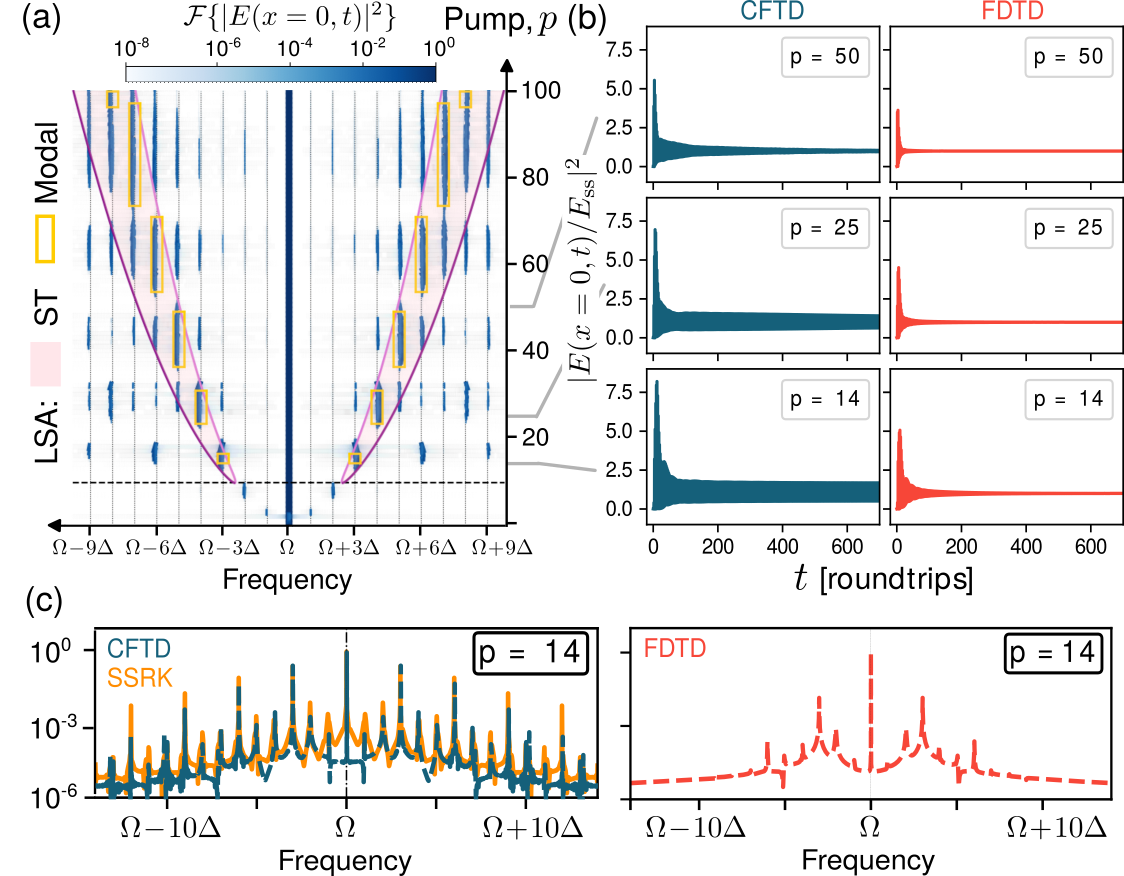}
    \caption[]{RNGH instability and comb formation for decay parameters: $\tbar{\gamma}_{\perp}=10.0,\tbar{\gamma}_{\parallel}=0.5,\tbar{\kappa}=0.1$, and $n_{\rm R}=1.96$. (a) Frequency spectrum of electric field as a function of pump power obtained using CFTD. Shaded regions mark the continuous frequencies predicted to be unstable by the ST LSA, while yellow boxes mark the discrete mode pairs predicted to be unstable by the modal LSA. The horizontal dashed line marks the RNGH instability threshold. (b) Side by side comparison of $|E(x=0,t)|^2$ scaled by $\ESS$ for the first 700 roundtrips of the CFTD (blue) and FDTD (red) results. The number of modes included in the CFTD simulations is $N = 19$. (c) Frequency spectrum of the electric field for $p = 14$ for CFTD (blue) and SSRK (orange) in the left panel, and FDTD (red) in the right panel, calculated for 1100 roundtrips. Note the qualitatively good agreement between the CFTD and the SSRK, and the significant deviations of FDTD.}
    \label{fig:above_thresh_g10}
\end{figure}

As the ratio of decay parameters increases, so does the number of relevant modes that must be included in the simulation. In this case, the number of modes increases up to $\sim$ 21 (corresponding to $\Omega \pm 10\Delta$). We then once again calculate the frequency spectrum of the electric field using CFTD, with the results shown in the left panel of Fig.~\ref{fig:above_thresh_g10}.

With increasing pump power beyond the RNGH threshold $p_{\rm th}$, a stable frequency comb emerges from the single-mode lasing state, as before. However the observed spacing is now equal to 3 FSRs, as the increased ratio of $\gamma_{\perp}/\gamma_{\parallel}$ increases the unstable mode frequency $\omega_{\rm inst}$ via Eq.~(\ref{eq:wmaxmin}). Importantly, the unstable mode pair is exactly that predicted by both LSA schemes. With further increase in pump power, the predicted unstable mode pair moves outwards relative to the single-mode lasing frequency. Importantly, the CFTD simulations show an increase in the comb FSR with pump power, consistent with both LSA methods. Furthermore, the observed combs can even possess multiple, non-commensurate dominant frequency components, unlike the simpler dynamics at lower values of $\gamma_{\perp}/\gamma_{\parallel}$.





In contrast, in these regimes our benchmark FDTD method performs rather poorly. In Fig.~\ref{fig:above_thresh_g10}(b), we compare simulated time traces using CFTD and FDTD for 700 roundtrips for three selected pump powers $p = 14$, $p = 25$ and $p = 50$. While CFTD demostrates sustained oscillations as necessitated by the comb spectra observed Fig.~\ref{fig:above_thresh_g10}(a), the FDTD simulation exhibits a suppressed amplitude and quickly settles to $E_{\rm ss}$. The difference is further highlighted by a detailed look at the frequency spectrum of the electric field at $p=14$ shown in Fig.~\ref{fig:above_thresh_g10}(c) for both CFTD and FDTD. The FDTD simulation clearly does not yield a broad frequency comb like the CFTD approach; rather the central mode and only two sidebands are the main active frequencies. The ``washing out'' of higher frequency components is a known issue of FDTD methods, due to the increased rate of accumulation of phase errors of higher frequency components, whose the phase evolves more rapidly. By explicitly operating in the frequency domain, the CFTD appears more robust to such effects. 
 
 


Deeper in this regime of fast polarization relaxation where increasing disagreements are observed between the CFTD and FDTD, the SSRK scheme provides a more exact method that can be used to understand the discrepancies. However, we recall at the outset that the SSRK is subject to a number of limitations. First, it is tailored to ring geometries only, whereas the CFTD can accommodate arbitrary geometries. Second, simulation times for the SSRK are similar to the FDTD; hence the CFTD remains a considerably faster method. Finally, the SSRK cannot be easily extended to simulate photonic molecule or other coupled laser arrangements. 
 
 With these caveats, we present comparisons of the CFTD against the SSRK in regimes of fast polarization relaxation, where the FDTD performs poorly. For ${\tbar{\gamma}_{\perp}}/{\tbar{\gamma}_{\parallel}}=20$, the SSRK generally shows good agreement with the CFTD. An example of the electric field frequency spectrum in the unstable regime is shown in orange in the lower panel of Fig.~\ref{fig:above_thresh_g10} revealing a 3-FSR comb for $p=14$; the prominent frequency peaks show good agreement with the corresponding spectrum using CFTD shown in blue in the left panel, and strongly disagrees with the FDTD in red in the right panel.
 

\begin{figure}[t]
    \centering
    \includegraphics[scale=0.48]{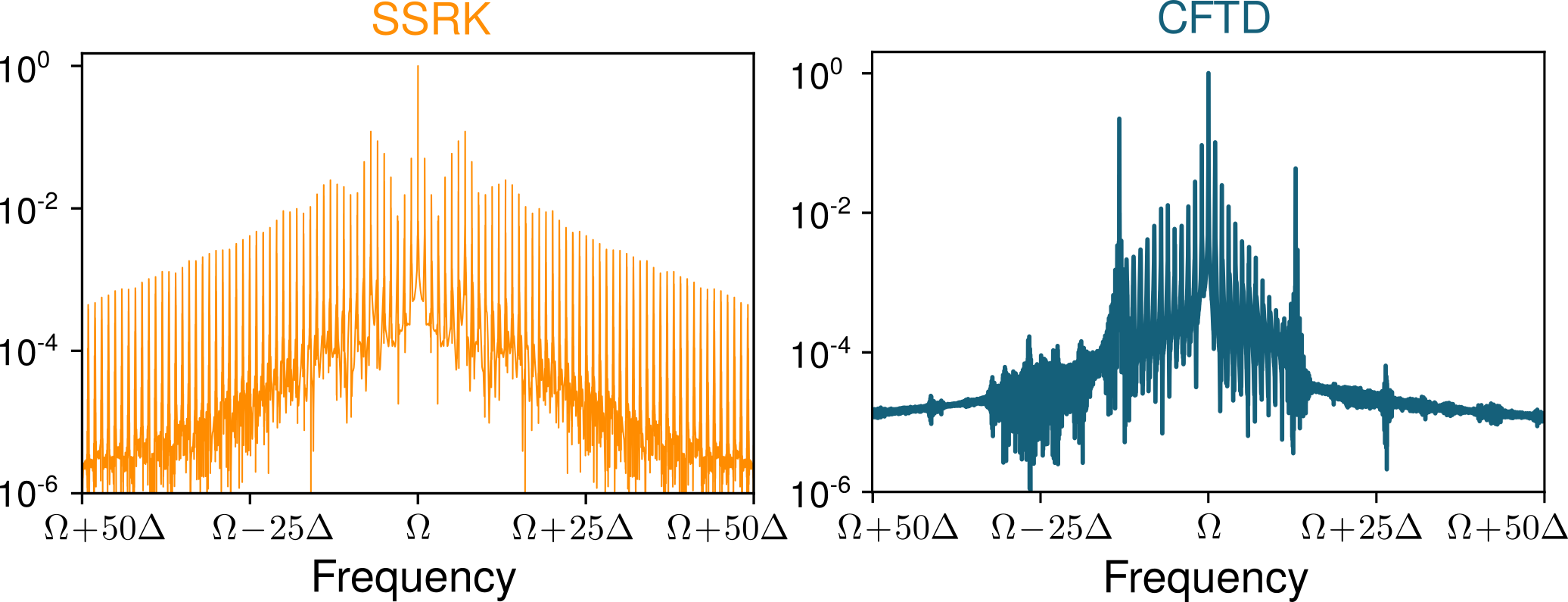}
    \caption[]{SSRK and CFTD comparison for decay parameters: $\tbar{\gamma}_{\perp}=25.0,\tbar{\gamma}_{\parallel}=0.5,\tbar{\kappa}=0.1$, and $n_{\rm R}=1.96$. The selected pump power is $p = 30$. Shown on the left is the frequency spectrum of the electric field as computed Fourier Transform from 1100 roundtrips of the SSRK result showing more than 100 modes. Shown on the right is the frequency spectrum of the electric field as obtained from 1000 roundtrips of the CFTD with $N=35$ modes included in the simulation.}
    \label{fig:SSM_CFTD}
\end{figure}


 
 
For even larger ${\tbar{\gamma}_{\perp}}/{\tbar{\gamma}_{\parallel}}$ ratios, we expect the CFTD to require the retention of even more spatial modes for simulation, leading to diminishing returns in terms of the relative simulation time advantage in comparison to the SSRK, and also possibly introducing numerical errors. We illustrate one such regime where ${\tbar{\gamma}_{\perp}}/{\tbar{\gamma}_{\parallel}}=50$ in Fig.~\ref{fig:SSM_CFTD}, showing the electric field frequency spectrum obtained using the SSRK method in orange and the CFTD in blue. We note that the SSRK predicts more than 100 distinct frequency components to be excited. In the CFTD simulation we include 35 spatial modes, which leads to a reasonable simulation time (shorter than for the SSRK) and no obvious numerical artefacts. While CFTD is not restricted to only exhibiting as many frequency components as included spatial modes (See Ref.~\cite{malik_spectral_2015, malik_nonlinear_2017} and discussion in Sec.~\ref{subsec:cftdDyn}), here we see the CFTD spectrum is significantly restricted in comparison to the SSRK. However, while CFTD may not be ideally suited to simulating such extremely broadband dynamics in ring lasers where the SSRK exists as an alternative, the generalizability to arbitrary geometries still makes CFTD an attractive proposition for analyzing multimode instabilities in more general laser systems.


\section{Chaotic dynamics}
\label{sec:chaos}





Not all multimode instabilities yield desired stable frequency combs. In this section, we will show that the CFTD can also efficiently capture a broader class of multimode instabilities, namely chaotic lasing dynamics. To this end, we now consider a different parameter regime to our prior simulations: we maintain $\tbar{\gamma}_{\parallel} = 0.5$, but set $\tbar{\gamma}_{\parallel}/\tbar{\gamma}_{\perp} = 1$, and importantly now choose $\tbar{\kappa} = 1.1$, which is just large enough for the laser system to operate in the ``bad cavity'' regime, where $\tbar{\kappa} > \tbar{\gamma}_{\perp}+\tbar{\gamma}_{\parallel}$. \sakc{Note that we again have $\gamma_{\perp} <\Delta$, the regime of the LH instability, which in conjunction with the bad cavity limit is known to feature chaotic emission at high pump powers~\cite{narducci_experimentally_1985, valcarcel_modal_2003}.} We now demonstrate that this is indeed the case, and analyze CFTD performance in such regimes.

\begin{figure}[t]
    \centering
    \includegraphics[scale=0.68]{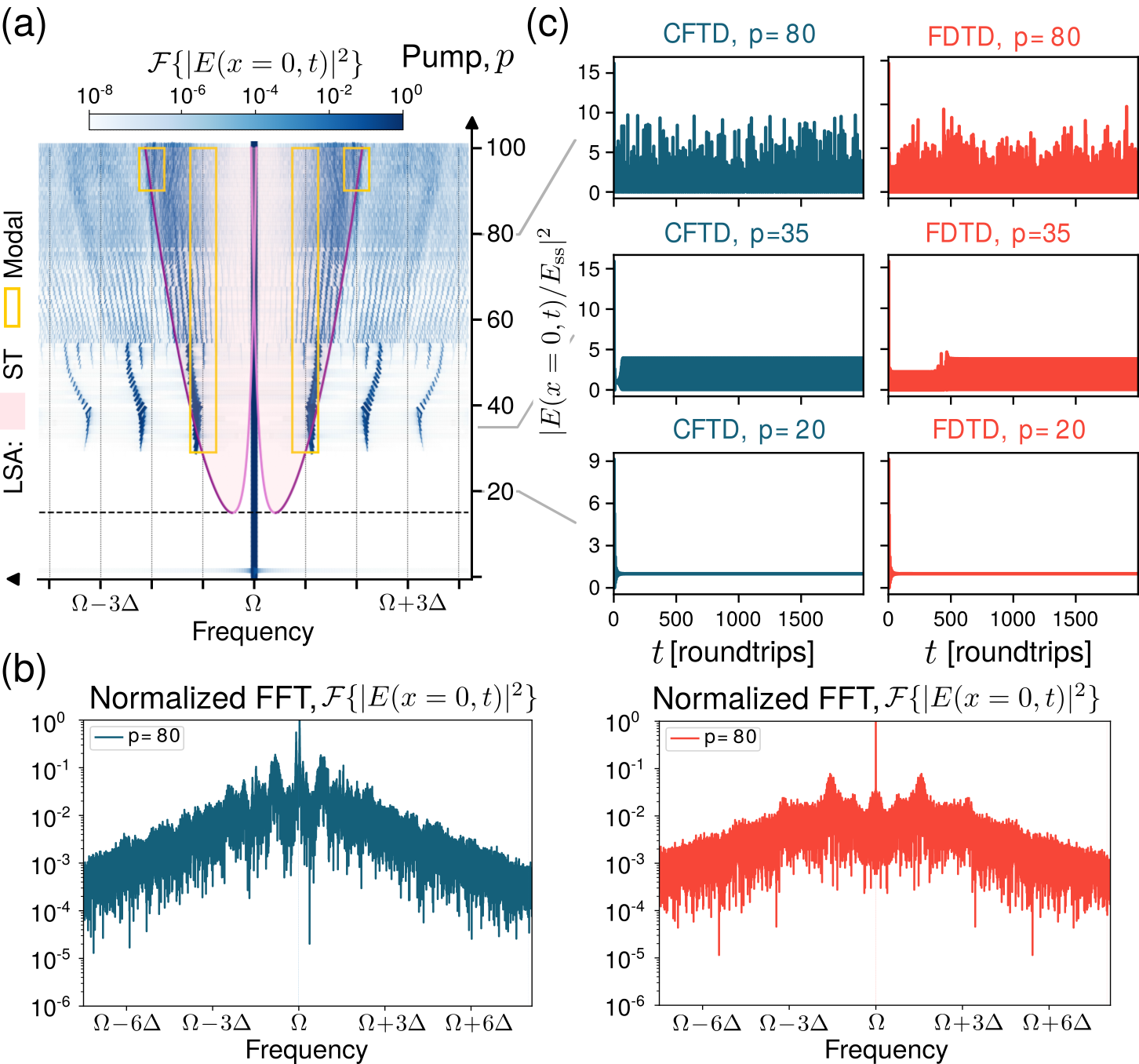}
    \caption[]{ Emergence of chaotic behavior for decay parameters $\tbar{\gamma}_{\perp}=0.5,\tbar{\gamma}_{\parallel}=0.5,\tbar{\kappa}=1.1$, and $n_{\rm R}=1.96$. (a) Frequency spectrum of the electric field as a function of pump power obtained using CFTD with $N=17$ retained modes. Shaded regions mark the continuous frequencies predicted to be unstable by the ST LSA, while yellow boxes mark the discrete mode pairs predicted to be unstable by the modal LSA. The horizontal dashed line indicates the instability threshold. (b) Side by side comparison of $|E(x=0,t)|^2$ scaled by $\ESS$ for the first 2000 roundtrips of the CFTD (blue) and FDTD (red) results. (c) Fourier Transform of $|E(x=0,t)|^2$ for $p = 80$ where we find chaotic behaviour in both the CFTD (blue) and FDTD (red) simulation. }
    \label{fig:chaotic}
\end{figure}

In Fig.~\ref{fig:chaotic} we show the spectrum of the electric field as a function of pump power using CFTD. For pump powers past $p_{\rm th}$ up to around $p \sim 30$, the single-mode lasing regime persists. This is consistent with the modal LSA, which predicts no unstable mode pairs in this pump range, and the ST LSA as no cold mode frequencies fall in the predicted instability region. For higher powers up to $p \sim 55$, a stable frequency comb solution emerges, initially with an FSR predicted by both LSAs, but exhibiting complex frequency pulling at higher pump powers. 

Such dynamics is similar to that observed in simulations in the good cavity regime in Sec.~\ref{sec:rngh}. However, beyond $p \sim 55$, the spectrum of the electric field suddenly broadens into several finely spaced, but still coherent, peaks. At even higher pump powers around $p \sim 70$, however, any sharp peaks decohere into an elevated and extremely broad noise floor, features typical of chaotic regimes. This qualitative change in lasing dynamics coincides with regions where the instability bands of the ST LSA, which are symmetric with respect to the central frequency, start to overlap. Such an overlap requires a system operating in the bad cavity regime~(see Appendix~\ref{app:rnghreview}), as we consider here. 

At a specific pump power $p=80$ in this regime, a more detailed look at the frequency spectrum of the electric field is shown in Fig.~\ref{fig:chaotic}(c), using both CFTD and FDTD schemes. Both show clear signatures of chaos: a broadband, noisy spectrum with several active frequencies and no significant coherent peaks. Time traces of the electric field using both CFTD and FDTD are also shown in Fig.~\ref{fig:chaotic}(b). For $p=80$, both methods are in good agreement, showing noisy traces with no recurrent oscillations. We also see that for specific pump powers in the other two aforemention regimes, CFTD and FDTD also show very good agreement.

We therefore clearly see that the CFTD is able to capture chaotic dynamics, as well as transitions from stable comb formation regimes to chaotic regimes. This makes the CFTD approach promising to identify - and avoid - regions of parameter space that do exhibit multimode instabilities, but do not allow useful, stable frequency comb formation.

\section{Simulation time comparison}
\label{sec:speedup}

\begin{figure}[t]
    \centering
    \includegraphics[scale=1]{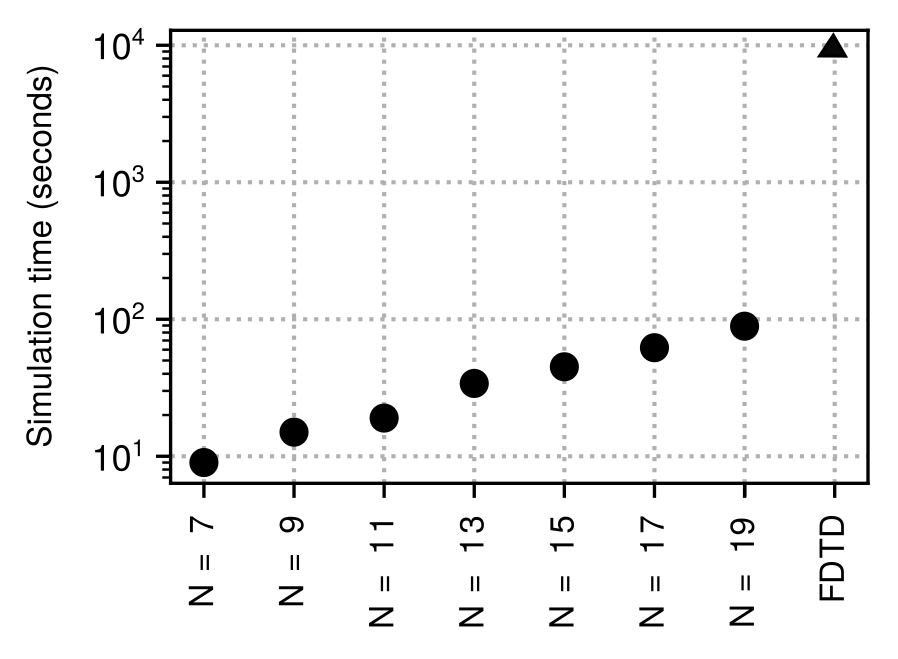}
    \caption{Simulation time comparison between the CFTD method for various numbers of modes included in the simulation (circles) and the FDTD method (triangle). The CFTD is at least two orders of magnitude faster, even when a relatively large number of modes is included in the simulation.}
    \label{fig:Simulation_time}
\end{figure}

Having explored the numerical fidelity of the CFTD in comparison to more standard simulation schemes, we now quantify the speedup in simulation time provided by the CFTD. In Fig.~\ref{fig:Simulation_time}, we show the simulation time for the CFTD as a function of the number of included modes $N$, and the simulation time for an equivalent FDTD integration for comparison. The results shown are for the laser system defined by parameters $\tbar{\gamma}_{\perp}=5.0,\tbar{\gamma}_{\parallel}=0.5,\tbar{\kappa}=0.1$, $n_{\rm R}=1.96$, and a fixed pump power $p=25$; however they are representative of the speedup typically observed using the CFTD. We note that to ensure a fair comparison, the number of roundtrips simulated and the spatial discretization used is kept the same between both methods. 

We consider various mode numbers from $N = 7$ to $N = 19$ for the CFTD, even though $N = 15$ is sufficient for the CFTD to yield accurate results in this particular example. The simulation time for the CFTD increases with the number of modes, but even for the highest mode numbers it remains about two orders of magnitude faster than the FDTD. This highlights one of the main advantages of our spectral CFTD approach: its computational speed, attained by the ability to reduce the integration of multidimensional PDEs to a set of ODEs in time only. The stark difference in the one-dimensional ring laser case explored here will only become larger for simulations of larger laser systems in two- or three-dimensions.


\section{Conclusions and Outlook}

In this paper, we have demonstrated a spectral approach, CFTD, as a viable and efficient method to simulate one of the most complex time-dependent phenomena in lasers: the emergence and resultant dynamics of coherent multimode instabilities. Our approach projects the standard MBEs of laser dynamics onto a suitably chosen spatial basis accounting for the lasing cavity geometry and losses, obtaining a set of time-dependent coefficients of the electric field, polarization, as well as inversion which is captured via a projected set of matrix elements. Expanding far beyond our previous work \cite{malik_spectral_2015}, we have considered regimes where non-stationary inversion plays a crucial role in the emergence of coherent multimode instabilities. By benchmarking the CFTD method with an FDTD approach, as well as an SSRK scheme specific to ring lasers, we have found excellent qualitative agreement across a very wide parameter space, and quantitative agreement in regimes with up to $\approx$ 30 relevant modes. We reveal the ability to capture various waveforms and complex phenomena in our analysis, starting from single mode behavior to frequency combs of different FSR spacings to broadband chaotic spectra. 

Our method provides not only an increase in simulation speed of at least two orders of magnitude over standard spatiotemporal integration schemes, but also more analytic insight into the system under study, via a linear stability analysis based on discrete included modes. While we have analyzed here a single ring laser, the CFTD method we introduce is a fast and efficient tool that can easily be extended to study complex time-dependent phenomena in much more general laser systems, a research area of significant contemporary interest. Such systems include complex cavity geometries without specific spatial symmetries, as well as coupled laser arrays. The complexity of analyzing such systems with conventional numerical schemes is substantial even in standard multimode lasing regimes (i.e. with stationary inversion), let alone coherent instabilities with nontrivial population dynamics. This only highlights the utility of the CFTD method.


\begin{acknowledgements}
This work was supported by the Eric and Wendy Schmidt Transformative Technology Fund. Simulations in this paper were performed using the Princeton Research Computing resources at Princeton University, which is a consortium of groups led by the Princeton Institute for Computational Science and Engineering (PICSciE) and Office of Information Technology's Research Computing.
\end{acknowledgements}

\appendix

\section{MBEs in the slowly-varying envelope approximation}

A standard approach to analyzing the dynamics of the MBEs of Eqs.~(\ref{eq:MBE})-(\ref{eq:MBD}) begins by extracting the fast frequency dependence from the atomic transition frequency:
\begin{subequations}
\begin{align}
    \mathcal{E}(x,t) &= E_c\cdot\tbar{E}(x,t) \frac{1}{\sqrt{L}} e^{i (n_{\rm R} \Omega /c) x} e^{-i\Omega t}  \\
    \mathcal{P}(x,t) &= P_c\cdot\tbar{P}(x,t) \frac{1}{\sqrt{L}} e^{i (n_{\rm R} \Omega /c) x} e^{-i\Omega t} \\
    \mathcal{D}(x,t) &= D_c\cdot \tbar{D}(x,t) 
\end{align}
\end{subequations}
Following the above ans\"atze, the primary transformation happens in the wave equation for the electric field, Eq.~(\ref{eq:MBE}):
\begin{align}
    &\partial_x^2 \tbar{E}+\frac{i2n_{\rm R}\Omega}{c}\partial_x \tbar{E}  +\frac{i2n^2\Omega}{c^2}\dot{\tbar{E}}+\frac{n^2-n_{\rm R}^2}{n^2}\frac{\Omega^2}{c^2}\tbar{E} - \frac{n^2}{c^2}\ddot{\tbar{E}} = \nonumber \\
    &{\mu}_0\epsilon_0\ddot{\tbar{P}} - i2\mu_0\epsilon_0 \Omega \dot{\tbar{P}} - \mu_0\epsilon_0\Omega^2 \tbar{P}
\end{align}
where we have used the explicit forms of $E_c$, $P_c$ as defined in Eq.~(\ref{eq:scalingF}). The slowly-varying envelope approximation relies on the fact that the spatio-temporal evolution of the envelope functions $\tbar{E}, \tbar{P}$ will unfold at rates smaller than $\Omega$ and $\Omega/c$. More precisely, this requires $(\Omega/c) \partial_x \tbar{E} \gg \partial_x^2 \tbar{E}$, $\Omega \dot{\tbar{E}} \gg \ddot{\tbar{E}}$, and with analogous conditions holding for the polarization field. Dropping these second-order derivatives, the electric field wave equation reduces to an advection equation of the form:
\begin{align}
    i \dot{\tbar{E}} = \left[-i\frac{n_{\rm R}}{n^2}c~\partial_x - \frac{n^2-n_{\rm R}^2}{n^2}\frac{\Omega}{2} \right]\tbar{E} - \frac{\Omega}{2n^2}\tbar{P}
\end{align}
where we have also used $\mu_0\epsilon_0 = \frac{1}{c^2}$. Now recalling that $n= n_{\rm R} + in_{\rm I}$ where $\frac{n_{\rm I}}{n_{\rm R}} \ll 1$, we can write the two respective contributions above to lowest nontrivial order in $\frac{n_{\rm I}}{n_{\rm R}}$:
\begin{align}
    \frac{n_{\rm R}}{n^2} &= \frac{1}{n_{\rm R}}\left(1 + i\frac{n_{\rm I}}{n_{\rm R}} \right)^{-2} \simeq \frac{1}{n_{\rm R}} - i\frac{2n_{\rm I}}{n_{\rm R}^2} \nonumber \\
    \frac{n^2-n^2_{\rm R}}{n^2} &= 1- \left(1 + i\frac{n_{\rm I}}{n_{\rm R}} \right)^{-2} \simeq  i\frac{2n_{\rm I}}{n_{\rm R}} 
\end{align}
Using these results allows us to write the advection equation to lowest nontrivial order in $\frac{n_{\rm I}}{n_{\rm R}}$,
\begin{align}
    i \dot{\tbar{E}} = \left[-i\frac{c}{n_{\rm R}}~\partial_x \tbar{E}-\frac{2n_{\rm I}}{n_{\rm R}^2}~c~\partial_x \tbar{E} - i\frac{2n_{\rm I}}{n_{\rm R}}\cdot\frac{\Omega}{2} \right]\tbar{E} - \frac{\Omega}{2n^2}\tbar{P} 
\end{align}
If we now define the cold cavity loss rate as
\begin{align}
    \kappa \equiv \left(\frac{n_{\rm I}}{n_{\rm R}} \right)\Omega,
\end{align}
we finally obtain:
\begin{align}
    \dot{\tbar{E}} = -\frac{c}{n_{\rm R}}\left( 1-i\frac{2n_{\rm I}}{n_{\rm R}} \right)\partial_x \tbar{E} - \kappa \tbar{E} + i \frac{\Omega}{2n_{\rm R}^2}\tbar{P} 
\end{align}
Finally, using Eq.~(\ref{eq:scalingV}), the above equation can be expressed in dimensionless form:
\begin{align}
    \dot{\tbar{E}} = -\frac{1}{n_{\rm R}}\left( 1-i\frac{2n_{\rm I}}{n_{\rm R}} \right)\partial_{\tbar{x}} \tbar{E} - \tbar{\kappa} \tbar{E} + i \frac{\tbar{\Omega}}{2n_{\rm R}^2}\tbar{P} 
\end{align}

Next, we substitute the ans\"atze into Eq.~(\ref{eq:MBP}) and simplify, which yields:
\begin{align}
\dot{\tbar{P}} = -\tbar{\gamma}_{\perp}\tbar{P} - i\tbar{\gamma}_{\perp}\tbar{E}\tbar{D}
\end{align}
Finally, substituting the ans\"atze into Eq.~(\ref{eq:MBD}) and simplifying using Eq.~(\ref{eq:scalingV}), we obtain:
\begin{align}
\dot{\tbar{D}} = -\tbar{\gamma}_{\parallel}(\tbar{D} - \tbar{D}^0) + i\frac{\tbar{\gamma}_{\parallel}}{2}(\tbar{E}\tbar{P}^* - \tbar{E}^*\tbar{P})
\end{align}

\section{Review of main results of RNGH LSA}
\label{app:rnghreview}

For completeness, we present here the results of the original spatiotemporal (ST) LSA from Ref.~\cite{risken_selfpulsing_1968}, in the notation of the main text. This LSA predicts minimum and maximum unstable frequencies at any pump power $p$ given respectively by:
\begin{align}
    \tbar{\omega}_{\rm max,min} &= \sqrt{ \frac{\tbar{\gamma}_{\parallel}}{2}\left(3\tbar{\gamma}_{\perp}(p-1)-\tbar{\gamma}_{\parallel}\pm R \right) }\times \nonumber \\
    &~~~~~\left(1 - \frac{2\tbar{\kappa} }{\tbar{\gamma}_{\perp}(p-3)-\tbar{\gamma}_{\parallel}\pm R} \right)
    \label{eq:wmaxmin}
\end{align}
where:
\begin{align}
    R = \sqrt{\tbar{\gamma}_{\perp}^2(p-1)^2 - 2\tbar{\gamma}_{\perp}(p-1)(4\tbar{\gamma}_{\perp}+3\tbar{\gamma}_{\parallel})+\tbar{\gamma}_{\parallel}^2 }  
\end{align}
We define as $\omega_{\rm inst}$ the value of the unstable frequency at the emergence of the instability of single mode lasing. Here $\omega_{\rm max}=\omega_{\rm min}$, and hence $R=0$. This also provides the value of the pump power threshold for the instability:
\begin{align}
    p_{\rm th} = 5 + 3\left(\frac{\tbar{\gamma}_{\parallel}}{\tbar{\gamma}_{\perp}} \right) + 2 \sqrt{4 + 6\left(\frac{\tbar{\gamma}_{\parallel}}{\tbar{\gamma}_{\perp}}\right) + 2\left(\frac{\tbar{\gamma}_{\parallel}}{\tbar{\gamma}_{\perp}}\right)^{\ 2} } 
\end{align}
To lowest order in $\frac{\tbar{\gamma}_{\parallel}}{\tbar{\gamma}_{\perp}}$, we recover the often-quoted value for the second thresold, $p_{\rm th} \to 9$. We note that this is only true in the large $\gamma_{\perp}$ limit.

Defining $\tbar{\Delta}_{\rm inst} = \tbar{\omega}_{\rm max}-\tbar{\omega}_{\rm min}$, considering the near threshold regime, we present results to lowest order in $\frac{\tbar{\kappa}}{\tbar{\gamma}_{\perp}},\frac{\tbar{\gamma}_{\parallel}}{\tbar{\gamma}_{\perp}}$ are presented in Eqs.~(\ref{eq:wmin}),~(\ref{eq:deltainst}).

The unstable frequency $\tbar{\omega}_{\rm inst}$ can become zero under specific circumstances. Requiring $\tbar{\omega}_{\rm inst} = 0$ using Eq.~(\ref{eq:wmaxmin}), we find that this condition is met at a pump power $p$ given by:
\begin{align}
    p(\tbar{\omega}_{\rm inst} = 0) = \frac{\tbar{\kappa}}{\tbar{\gamma}_{\perp}}\cdot\frac{\tbar{\gamma}_{\parallel}+3\tbar{\gamma}_{\perp}+\tbar{\kappa}  }{ \tbar{\kappa}-\tbar{\gamma}_{\parallel}-\tbar{\gamma}_{\perp} }
\end{align}
Clearly for this pump power to be positive and hence physical, we require:
\begin{align}
    \tbar{\kappa}> \tbar{\gamma}_{\parallel}+\tbar{\gamma}_{\perp}
\end{align}
which is the ``bad cavity'' limit of laser operation.

\section{Lossy modes of multimode ring cavity}
\label{app:ringCavityModes}

In this appendix section we will solve the Helmholtz equation, Eq.~(\ref{eq:helmholtz}), for the modes of a lossy multimode ring cavity. For convenience we reproduce the equation below, specialized to the one-dimensional ring cavity:
\begin{align}
    \frac{d^2}{dx^2} \varphi_m(x) = -\frac{n^2 \omega_m^2}{c^2}\varphi_m(x)
    \label{eq:modeDef}
\end{align}
We begin by writing the equation in dimensionless parameters, as defined by Eq.~(\ref{eq:scalingV}),
\begin{align}
    \frac{d^2}{d\tbar{x}^2} \varphi_m(\tbar{x}) = -n^2\tbar{\omega}_m^2\varphi_m(\tbar{x})
\end{align}
The above Helmholtz equation can be solved by assuming ans\"atze for the mode functions $\varphi_m(\tbar{x})$ of the form:
\begin{align}
    \varphi_m(\tbar{x}) = N e^{i\tbar{k}_m\tbar{x}}
    \label{eq:anshelmholtz}
\end{align}
where $\tbar{k}_m$ is a dimensionless wavevector, related to the dimensionful wavevector $k_m$ via $\tbar{k}_m = Lk_m$, and where $N$ is a normalization constant. Note that for Eq.~(\ref{eq:anshelmholtz}) to satisfy the Helmholtz equation for the ring cavity, it must satisfy periodic boundary conditions, so that:
\begin{align}
    e^{i\tbar{k}_m\tbar{x}} = e^{i\tbar{k}_m(\tbar{x}+1)} \implies \tbar{k}_m = 2m\pi,~m\in\mathbb{Z} 
\end{align}
Substituting the above and simplifying, we obtain the relation:
\begin{align}
   \left[  -\tbar{k}_m^2 + n^2\tbar{\omega}_m^2 \right]\varphi_m(\tbar{x}) = 0
\end{align}
The above immediately yields the dispersion relation:
\begin{align}
    \tbar{\omega}_m = \frac{1}{n}\tbar{k}_m 
\end{align}
Recalling that $n$ is complex-valued, the eigenfrequencies $\omega_m$ must also be complex, and describe the decay of field intensity for the lossy ring cavity modes (in the absence of pumping and subsequent lasing).
\begin{align}
    \tbar{\omega}_m &= \frac{1}{n_{\rm R}+in_{\rm I}}\tbar{k}_m \nonumber \\
    &\simeq \frac{k_m}{n_{\rm R}}- i \frac{\tbar{k}_m}{n_{\rm R}}\frac{n_{\rm I}}{n_{\rm R}} \equiv \tbar{\nu}_m -i\tbar{\kappa}_m
    \label{appeq:omegam}
\end{align}
Recalling that we have chosen parameters such that $\tbar{\nu}_0 = \tbar{\Omega}$, we find:
\begin{align}
    \tbar{\nu}_0 = \frac{\tbar{k}_0}{n_{\rm R}} = \tbar{\Omega},~\tbar{\kappa}_0 = \frac{\tbar{k}_0}{n_{\rm R}}\frac{n_{\rm I}}{n_{\rm R}} = \tbar{\Omega}\frac{n_{\rm I}}{n_{\rm R}} = \tbar{\kappa}
\end{align}
Finally, imposing the orthogonality relationship in Eq.~(\ref{eq:orth}), we find that $N= \frac{1}{\sqrt{L}}$, so that the final expression for lossy ring cavity modes takes the form:
\begin{align}
    \varphi_m(x) = \frac{1}{\sqrt{L}} e^{ik_mx},~k_mL = 2m\pi,~m\in\mathbb{Z} 
\end{align}
with complex eigenfrequencies given by Eq.~(\ref{appeq:omegam}).

\section{Derivation of CFTD equations}
\label{app:cftd}

We will derive here Eq.~(\ref{eq:CFTD_set}) starting with the Maxwell-Bloch equations for the scalar electric field amplitude $E(x,t)$, polarization $P(x,t)$ and inversion density $D(x,t)$, Eqs.~(\ref{eq:MBE})-(\ref{eq:MBD}). The procedure for obtaining the CFTD equations can be summarized as follows: we substitute the CF ans\"atze, Eqs.~(\ref{eq:ansE}),~(\ref{eq:ansP}) into each of the Maxwell-Bloch equations, integrate out the spatial component by taking advantage of properties of the CF states, and apply the slowly varying envelope approximation to eliminate second-order time derivatives. 

Beginning by substituting Eqs.~(\ref{eq:ansE}),~(\ref{eq:ansP}) into Eq.~(\ref{eq:MBE}), we obtain:
\begin{align}
    &\sum_m\left[ \frac{\tbar{E}_m}{E_c}{\nabla}^2  -\frac{n^2}{c^2}\frac{\ddot{\tbar{E}}_m-i2\Omega\dot{\tbar{E}}_m -\Omega^2 \tbar{E}_m  }{E_c}\right] e^{-i\Omega t}\varphi_m(x) \nonumber \\
    &= {\mu}_0\sum_m\frac{\ddot{\tbar{P}}_m-i2\Omega\dot{\tbar{P}}_m -\Omega^2 \tbar{P}_m  }{P_c} e^{-i\Omega t}\varphi_m(x)
\end{align}
Using Eq.~(\ref{eq:modeDef}) the above equation simplifies to:
\begin{align}
    &\sum_m\left[ -\frac{n^2}{c^2}\frac{\omega_m^2 \tbar{E}_m}{E_c} -\frac{n^2}{c^2}\frac{\ddot{\tbar{E}}_m-i2\Omega\dot{\tbar{E}}_m -\Omega^2 \tbar{E}_m  }{E_c}\right] e^{-i\Omega t}\varphi_m(x) \nonumber \\
    &= {\mu}_0\sum_m\frac{\ddot{\tbar{P}}_m-i2\Omega\dot{\tbar{P}}_m -\Omega^2 \tbar{P}_m  }{P_c} e^{-i\Omega t}\varphi_m(x)
\end{align}
Multiplying through by $\varphi_n(x)$ and integrating over the spatial domain of the cavity, we find:
\begin{align}
        &\sum_m\left[ -\frac{n^2}{c^2}\frac{\omega_m^2 \tbar{E}_m}{E_c} -\frac{n^2}{c^2}\frac{\ddot{\tbar{E}}_m-i2\Omega\dot{\tbar{E}}_m -\Omega^2 \tbar{E}_m  }{E_c}\right] e^{-i\Omega t}\times \nonumber \\
        &~~~~\int dx~\varphi_n(x)\varphi_m(x) \nonumber \\
    &= {\mu}_0\sum_m\frac{\ddot{\tbar{P}}_m-i2\Omega\dot{\tbar{P}}_m -\Omega^2 \tbar{P}_m  }{P_c} e^{-i\Omega t}\int dx~\varphi_n(x)\varphi_m(x)
\end{align}
This spatial projection enables us to make use of the orthogonality of the cavity modes $\{\varphi_m(x)\}$, Eq.~(\ref{eq:orth}), which collapses the sum over modes, finally yielding: 
\begin{align}
        &\left[ -\frac{n^2}{c^2}\frac{\omega_m^2 \tbar{E}_m}{E_c} -\frac{n^2}{c^2}\frac{\ddot{\tbar{E}}_m-i2\Omega\dot{\tbar{E}}_m -\Omega^2 \tbar{E}_m  }{E_c}\right] e^{-i\Omega t} \nonumber \\
    &= {\mu}_0\frac{\ddot{\tbar{P}}_m-i2\Omega\dot{\tbar{P}}_m -\Omega^2 \tbar{P}_m  }{P_c} e^{-i\Omega t}
\end{align}
We now also perform a slowly-varying envelope approximation to neglect second-order time derivatives as before, namely $\ddot{\tbar{E}}_m\ll \Omega\dot{\tbar{E}}_m$ and $\ddot{\tbar{P}}_m \ll \Omega\dot{P}_m \ll \Omega^2 \tbar{P}_m$. Also making use of the explicit form of scaling factors $E_c$, $P_c$ and rearranging terms, we arrive at:
\begin{align}
-i2\Omega \dot{\tbar{E}}_m - \Omega^2 \tbar{E}_m + \omega_m^2 \tbar{E}_m = \frac{\Omega^2\mu_0\epsilon_0 c^2}{n^2} {\tbar{P}}_m
\end{align}
We will now introduce some replacements and a final set of approximations. Using $\mu_0\epsilon_0 = \frac{1}{c^2}$, $\omega_m^2 = \nu_m^2-\kappa_m^2 - 2i\nu_m\kappa_m$, and approximating $\frac{\nu_m}{\Omega} \simeq 1$, $n \simeq n_{\rm R}(1 + i\frac{n_{\rm I}}{n_{\rm R}} ) \simeq n_{\rm R}$, and scaling time and frequency scales according to Eq.~(\ref{eq:scalingV}), we arrive at the dynamical equation for the electric field amplitude of the $n$th mode is given by:
\begin{align}
\dot{\tbar{E}}_m &= \frac{i}{2\tbar{\Omega}}\left(\tbar{\Omega}^2 - \tbar{\nu}_m^2 + \tbar{\kappa}_m^2 \right)\tbar{E}_m - \tbar{\kappa}_m\tbar{E}_m +\frac{i\tbar{\Omega}}{2n_{\rm R}^2}\tbar{P}_m
\end{align}
Which is Eq.~(\ref{eq:ET}) from the main text.

Similarly substituting Eqs.~(\ref{eq:ansE}),~(\ref{eq:ansP}), into the dynamical equation for the polarization field, Eq.~(\ref{eq:MBP}), we obtain:
\begin{align}
    &\sum_m \left[-i\Omega\tbar{P}_m + \dot{\tbar{P}}_m \right]\varphi_m(x) = \sum_m (-i\Omega - \gamma_{\perp} )\tbar{P}_m\varphi_m(x) \nonumber \\
    & -i\frac{\varg^2}{\hbar}\frac{P_c}{E_cD_c} \sum_m \tbar{E}_m \tbar{D}(x,t)\varphi_m(x)
\end{align}
Multiplying through by $\varphi_n(x)$ and integrating over the spatial domain of the cavity, we find once more:
\begin{align}
    \dot{\tbar{P}}_m  =  - \gamma_{\perp} \tbar{P}_m -i\gamma_{\perp}\sum_m \tbar{E}_m \varphi_n^*(x)\tbar{D}(x,t)\varphi_m(x)
\end{align}
where we have also used the explicit forms of $E_c$, $P_c$, and $D_c$. The term in brackets on the right defines the inversion matrix elements introduced in Eq.~(\ref{eq:ansD}). Introducing these matrix elements, the equation of motion for the polarization field expansion coefficients is given by:
\begin{align}
\dot{\tbar{P}}_n &= -\tbar{\gamma}_{\perp} \tbar{P}_n - i\tbar{\gamma}_{\perp} \sum_m \tbar{E}_m\tbar{D}_{nm} 
\end{align}

The equations of motion for the inversion matrix elements $\tbar{D}_{nm}$ are given by:
\begin{align}
    \dot{\tbar{D}}_{nm} &= -\tbar{\gamma}_{\parallel}(\tbar{D}_{nm}-\tbar{D}^0_{nm})  \nonumber \\
    &~~~~+\frac{i\tbar{\gamma}_{\parallel}}{2}\sum_{r}\!\left[\tbar{E}_{r}\tbar{P}_{n+r-m}^* - \tbar{E}_{n+r-m}^*\tbar{P}_{r}\right] 
\end{align}

Finally, substituting Eqs.~(\ref{eq:ansE}),~(\ref{eq:ansP}), into the dynamical equation for the inversion field, Eq.~(\ref{eq:MBP}), we obtain:
\begin{align}
    \dot{D} &= -\gamma_\parallel(D - D^0) \nonumber \\
    &~~~+ i\frac{2}{\hbar E_cP_c}\sum_{rs}\left[\tbar{E}_r\tbar{P}_s^*\varphi_r(x)\varphi_s^*(x) - \tbar{E}_r^*\tbar{P}_s\varphi_r^*(x)\varphi_s(x)\right]
\end{align}
Multiplying through by the scaling factor $D_c$, and re-ordering the sum by a redefinition of labels, we obtain:
\begin{align}
    \dot{\tbar{D}} &= -\gamma_\parallel(\tbar{D} - \tbar{D}^0) \nonumber \\
    &~~~+ i\frac{2D_c}{\hbar E_cP_c}\sum_{rs}(\tbar{E}_r\tbar{P}_s^* - \tbar{E}_s^*\tbar{P}_r)\varphi_r(x)\varphi_s^*(x)
\end{align}
We can project the inversion field onto the spatial basis. We multiply through by $\varphi_n(x)\varphi_m^*(x)$ and integrate over the spatial domain of the laser cavity, finally arriving at:
\begin{align}
    \dot{\tbar{D}}_{nm} &= -\tbar{\gamma}_\parallel(\tbar{D}_{nm} - \tbar{D}_{nm}^0) + i\frac{\tbar{\gamma}_{\parallel}}{2}\sum_{rs}\tbar{\mathcal{A}}_{nmrs}(\tbar{E}_r\tbar{P}_s^* - \tbar{E}_r^*\tbar{P}_s)
\end{align}
where we have used the explicit forms of the scaling factors, moved to dimensionless time and frequency variables, and introduced the dimensionless mode overlap tensor of the main text, Eq.~(\ref{eq:Adef}):
\begin{align}
    \tbar{\mathcal{A}}_{nmrs} = L\int_{0}^{L} dr~\varphi_n\varphi_m^*\varphi_r\varphi_s^*
\end{align}

\section{Derivation of the single mode threshold}
\label{app:sml}

In this appendix section, we derive the pump threshold and steady-state lasing solution in the regime of single-mode lasing. We allow for a general lasing frequency $\Omega_l$:
\begin{align}
    \tbar{E}_l(t) &= \ESS e^{-i\tbar{\Omega}_l t} \nonumber \\
    \tbar{P}_l(t) &= \PSS e^{-i\tbar{\Omega}_l t} \nonumber \\
    \tbar{D}_{ll}(t) &= \DSS
\end{align}
where the inversion is stationary. Under this ans\"atz, the modal equations then simplify to:
\begin{subequations}
\label{eq:singlemode}
\begin{align}
   \dot{\tbar{E}}_{\rm ss} &= \frac{i}{2\tbar{\Omega}}\left(2\tbar{\Omega}\tbar{\Omega}_l + \tbar{\Omega}^2 - \tbar{\nu}_l^2 + \kappa_l^2 \right)\ESS - \kappa_{0}\ESS +\frac{i\tbar{\Omega}}{2n_{\rm R}^2}\PSS \label{eq:ET_singlemode}  \\
    \dot{\tbar{P}}_{\rm ss} &= (i\tbar{\Omega}_l-\gamma_{\perp}) \PSS - i\gamma_{\perp} \ESS\DSS \label{eq:PT_singlemode} \\
    \dot{\tbar{D}}_{\rm ss} &= -\gamma_{\parallel}(\DSS-\tbar{D}^0_{00}) + \frac{i\gamma_{\parallel}}{2}(\ESS\PSS^* - \ESS^*\PSS) \label{eq:DT_singlemode}    
\end{align}
\end{subequations}
For steady-state, we require $\dot{\tbar{E}}_{\rm ss} = \dot{\tbar{P}}_{\rm ss} = \dot{\tbar{D}}_{\rm ss} = 0$. In this regime, Eq.~(\ref{eq:ET_singlemode}) reduces to:
\begin{align}
    \PSS = \left[ -i\frac{2 n_{\rm R}^2 \tbar{\kappa}}{\tbar{\Omega}} - \frac{n_{\rm R}^2}{\tbar{\Omega}^2}\left(2\tbar{\Omega}\tbar{\Omega}_l + \tbar{\Omega}^2 - \tbar{\nu}_l^2 + \tbar{\kappa}_l^2 \right) \right]\ESS
\end{align}
Similarly, Eq.~(\ref{eq:PT_singlemode}) yields for the same quantity:
\begin{align}
    \PSS = \frac{i\gamma_{\perp}}{i\tbar{\Omega}_l-\gamma_{\perp}}\cdot \ESS\DSS
\end{align}
Comparing the two equations, we immediately find for $\DSS$:
\begin{align}
    \frac{i\gamma_{\perp}}{i\tbar{\Omega}_l-\gamma_{\perp}}\cdot \DSS = \left[ -i\frac{2 n_{\rm R}^2 \tbar{\kappa}_l}{\tbar{\Omega}} - \frac{n_{\rm R}^2}{\tbar{\Omega}^2}\left(2\tbar{\Omega}\tbar{\Omega}_l + \tbar{\Omega}^2 - \tbar{\nu}_l^2 + \tbar{\kappa}_l^2 \right) \right]
\end{align}
Or, rationalizing the left-hand side:
\begin{align}
    \frac{\tbar{\Omega}_l\gamma_{\perp}-i\gamma_{\perp}^2}{\tbar{\Omega}^2_l+\gamma_{\perp}^2}\DSS = \left[ -i\frac{2 n_{\rm R}^2 \tbar{\kappa}_l}{\tbar{\Omega}} - \frac{n_{\rm R}^2}{\tbar{\Omega}^2}\left(2\tbar{\Omega}\tbar{\Omega}_l + \tbar{\Omega}^2 - \tbar{\nu}_l^2 + \tbar{\kappa}_l^2 \right) \right]
\end{align}
Comparing both sides, the imaginary parts immediately yield:
\begin{align}
    \DSS = \frac{\tbar{\Omega}^2_l+\gamma_{\perp}^2}{\gamma_{\perp}^2}\frac{2n_{\rm R}^2\tbar{\kappa}_l}{\tbar{\Omega}}
    \label{eq:DSS}
\end{align}
while the real parts yield:
\begin{align}
    \frac{\tbar{\Omega}_l\gamma_{\perp}}{\tbar{\Omega}^2_l+\gamma_{\perp}^2}\DSS = - \left[ \frac{n_{\rm R}^2}{\tbar{\Omega}^2}\left(2\tbar{\Omega}\tbar{\Omega}_l + \tbar{\Omega}^2 - \tbar{\nu}_l^2 + \tbar{\kappa}_l^2 \right) \right]
\end{align}
Substituting the expression for $\DSS$ into the above, we immediately find:
\begin{align}
    \frac{\tbar{\Omega}_l}{\gamma_{\perp}}\frac{2n_{\rm R}^2\tbar{\kappa}_l}{\tbar{\Omega}} = - \left[ \frac{n_{\rm R}^2}{\tbar{\Omega}^2}\left(2\tbar{\Omega}\tbar{\Omega}_l + \tbar{\Omega}^2 - \tbar{\nu}_l^2 + \tbar{\kappa}_l^2 \right) \right]
\end{align}
which finally yields:
\begin{align}
    \frac{2n_{\rm R}^2}{\tbar{\Omega}}\left( \frac{\tbar{\kappa}_l}{\gamma_{\perp}} +  1 \right)\tbar{\Omega}_l = -  \frac{n_{\rm R}^2}{\tbar{\Omega}^2}\left(\tbar{\Omega}^2 - \tbar{\nu}_l^2 + \tbar{\kappa}_l^2 \right) 
\end{align}
or:
\begin{align}
    \left( \frac{\tbar{\kappa}_l}{\gamma_{\perp}} +  1 \right)\tbar{\Omega}_l = -  \frac{\left(\tbar{\Omega}^2 - \tbar{\nu}_l^2 + \tbar{\kappa}_l^2 \right)}{\tbar{\Omega}} 
    \label{eq:OmegaL}
\end{align}

Finally, using Eq.~(\ref{eq:DT_singlemode}) we find:
\begin{align}
    0 = -\gamma_{\parallel}(\DSS-\tbar{D}^0_{00}) + \frac{i\gamma_{\parallel}}{2}\!\left(  \frac{i\gamma_{\perp}}{i\tbar{\Omega}_l+\gamma_{\perp}} -  \frac{i\gamma_{\perp}}{i\tbar{\Omega}_l-\gamma_{\perp}}   \right)\! \DSS |\ESS|^2
\end{align}
which simplifies to:
\begin{align}
    \DSS &= \tbar{D}^0_{00} - \left( \frac{\gamma_{\perp}^2}{\tbar{\Omega}^2_l+\gamma_{\perp}^2} \right) \DSS |\ESS|^2 \nonumber \\
    \implies \DSS &= \frac{\tbar{D}^0_{00}}{1 + \frac{\gamma_{\perp}^2}{\tbar{\Omega}^2_l+\gamma_{\perp}^2}|\ESS|^2 }
\end{align}
Using Eq.~(\ref{eq:DSS}), we can solve for $\ESS$:
\begin{align}
     \left( \frac{\tbar{\Omega}^2_l+\gamma_{\perp}^2}{\gamma_{\perp}^2} + |\ESS|^2 \right)\frac{2n_{\rm R}^2\tbar{\kappa}_l}{\tbar{\Omega}} = \tbar{D}^0_{00}
\end{align}
which can be rewritten in the form:
\begin{align}
    |\ESS|^2 = \frac{\tbar{\Omega}}{2n_{\rm R}^2\tbar{\kappa}_l}\tbar{D}^0_{00} - \frac{\tbar{\Omega}^2_l+\gamma_{\perp}^2}{\gamma_{\perp}^2}
\end{align}
The requirement of $|\ESS|^2>0$ yields:
\begin{align}
    \tbar{D}^0_{00} \geq \frac{2n_{\rm R}^2\kappa_l}{\tbar{\Omega}}\cdot\frac{\tbar{\Omega}^2_l+\gamma_{\perp}^2}{\gamma_{\perp}^2}
\end{align}
which yields the single-mode lasing threshold $\DthT$:
\begin{align}
    \DthT = \frac{2n_{\rm R}^2\tbar{\kappa}_l}{\tbar{\Omega}}\cdot\frac{\tbar{\Omega}^2_l+\gamma_{\perp}^2}{\gamma_{\perp}^2}
\end{align}
Clearly minimizing this threshold for fixed damping parameters requires minimizing $\tbar{\kappa}_l$ and $\tbar{\Omega}_l$. Assuming all modes have equal damping rates $\kappa_l = \kappa~\forall~l$, the lasing threshold is minimized for mode $l$ that minimizes $\tbar{\Omega}_l$. From Eq.~(\ref{eq:OmegaL}), $\tbar{\Omega}_l = 0$ if $\left(\tbar{\Omega}^2 - \tbar{\nu}_l^2 + \tbar{\kappa}_l^2 \right) = 0$, which requires:
\begin{align}
    \tbar{\nu}_l^2 = \left(\tbar{\Omega}^2 + \tbar{\kappa}_l^2 \right) \implies \tbar{\nu}_l \simeq \tbar{\Omega}
\end{align}
as $\tbar{\kappa}_l \ll \tbar{\Omega}$. Namely, the mode $l$ closest to the atomic transition frequency will have the lowest single-mode lasing threshold.





\section{Details of the multimode LSA}
\label{app:stability}

To analyze the linear stability of single-mode lasing dynamics described by the CFTD equations, we substitute the ans\"atze of Eqs.~(\ref{eq:linAnsatze}) into Eqs.~(\ref{eq:ET})-(\ref{eq:DT}), and retain only terms linear in the fluctuation variables $(\delta \tbar{E}_m,\delta \tbar{P}_m, \delta \tbar{D}_{nm})$. Under this linearization, the equation of motion for the electric field amplitudes, which is already linear, retains its form and is given by Eq.~(\ref{eq:ET}):
\begin{align}
    \delta\dot{\tbar{E}}_m &= \frac{i}{2\tbar{\Omega}}\left(\tbar{\Omega}^2 - \tbar{\nu}_{m}^2 + \tbar{\kappa}_{m}^2 \right)\delta\tbar{E}_{m} - \tbar{\kappa}_{m}\delta\tbar{E}_m +\frac{i\tbar{\Omega}}{2n_{\rm R}^2}\delta\tbar{P}_m \label{eq:dET}
\end{align}
while the equation of motion for polarization field, Eq.~(\ref{eq:PT}), for sidebands $m \neq 0$ take the form:
\begin{align}
    \delta\dot{\tbar{P}}_m &= -\tbar{\gamma}_{\perp} \delta\tbar{P}_m - i\tbar{\gamma}_{\perp} \!\! \left[ \tbar{E}_{0}\delta\tbar{D}_{m0} + \sum_{l\neq 0} \delta\tbar{E}_{l}\tbar{D}_{00}\delta_{ml} \right]
    \label{eq:dPT}
\end{align}
where we have dropped terms involving the product of two fluctuation terms $\sim O(\delta\tbar{E}_{m}\delta\tbar{D}_{ml}), m \neq 0$ in the second line. It is clear that the nontrivial inversion matrix elements that couple to the emerging sidebands are off-diagonal elements namely $\tbar{D}_{m0}$. Using Eq.~(\ref{eq:DT}), we can write equations of motion for these inversion matrix elements:
\begin{align}
    \delta\dot{\tbar{D}}_{m0} &= -\tbar{\gamma}_{\parallel}(\delta\tbar{D}_{m0}-\tbar{D}^0_{m0})  \nonumber \\
    &~~~~+\frac{i\tbar{\gamma}_{\parallel}}{2}\!\left[\tbar{E}_{0}\delta\tbar{P}_{m}^* + \delta\tbar{E}_{-m}\tbar{P}_{0}^* - \delta\tbar{E}_{m}^*\tbar{P}_{0} - \tbar{E}_{0}^*\delta\tbar{P}_{-m}\right] 
    \label{eq:dDT}
\end{align}
where we have now dropped second-order terms such as $O(\delta\tbar{E}_{r}\delta\tbar{P}_{m+r}),r\neq 0, -m$.

Eqs.~(\ref{eq:dET})-(\ref{eq:dDT}) can be compactly written in matrix form, yielding the matrix system in Eq.~(\ref{eq:jacobian}) of the main text. The component matrices are given by:
\begin{widetext}
\begin{align}
    \mathbf{J}_m = 
    \begin{pmatrix}
    \frac{i(\Omega^2+\kappa_m^2-\nu_m^2) - 2\kappa_m \Omega}{\tbar{\Omega}} & 0 & \frac{i\Omega}{2n_c^2} & 0 & 0 & 0 \\
    0 & \frac{i(\Omega^2+\kappa_m^2-\nu_m^2) - 2\kappa_m \Omega}{\tbar{\Omega}} & -\frac{i\Omega}{2n_c^2} & 0 & 0 \\
    - i \tbar{D}_{00}\tbar{\gamma}_{\perp} & 0 & -\tbar{\gamma}_{\perp} & 0 & -i \tbar{E}_{0}\tbar{\gamma}_{\perp} & 0 \\
    0 & i\tbar{D}_{00}\tbar{\gamma}_{\perp} & 0  & -\tbar{\gamma}_{\perp} & 0 & i \tbar{E}^*_{0}\tbar{\gamma}_{\perp} \\
    \frac{i}{2}\tbar{\gamma}_{\parallel}\tbar{P}_0^* & 0 & -\frac{i}{2}\tbar{\gamma}_{\parallel}\tbar{E}_0^* & 0 & -\tbar{\gamma}_{\parallel} & 0 \\
    0 & -\frac{i}{2}\tbar{\gamma}_{\parallel}\tbar{P}_0 & 0 & \frac{i}{2}\tbar{\gamma}_{\parallel}\tbar{E}_0 & 0 & -\tbar{\gamma}_{\parallel}
    \end{pmatrix}
\end{align}
\end{widetext}
The dependence on mode index $m$ arises only via the electric field sector. The coupling matrices $\mathbf{C}_{m,-m}$ are symmetric under $m\to-m$ are given by:
\begin{align}
    \mathbf{C}_{m,-m} = 
    \begin{pmatrix}
    0 & 0 & 0 & 0 & 0 & 0 \\
    0 & 0 & 0 & 0 & 0 & 0 \\
    0 & 0 & 0 & 0 & 0 & 0 \\
    0 & 0 & 0 & 0 & 0 & 0 \\
    0 & -\frac{i}{2}\tbar{\gamma}_{\parallel}\tbar{P}_0 & 0 & \frac{i}{2}\tbar{\gamma}_{\parallel}\tbar{E}_0 & 0 & 0 \\
    \frac{i}{2}\tbar{\gamma}_{\parallel}\tbar{P}^*_0 & 0 & -\frac{i}{2}\tbar{\gamma}_{\parallel}\tbar{E}^*_0 & 0 & 0 & 0 \\
    \end{pmatrix}
\end{align}
The coupling matrix has a sparse form, and clearly emphasizes that sideband coupling arises via the inversion field sector of the dynamics (the final two rows).



\section{Additional numerical results}
\label{app:simTime}

In this section we provide some additional simulation results that supplement discussions in the main text regarding the comparison of the two methods as well as other features of the CFTD.

\subsection{An intermediate polarization regime: $\gamma_{\perp}/\gamma_{\parallel} = 10.0$}

\begin{figure}[t]
    \centering
    \includegraphics[scale=0.83]{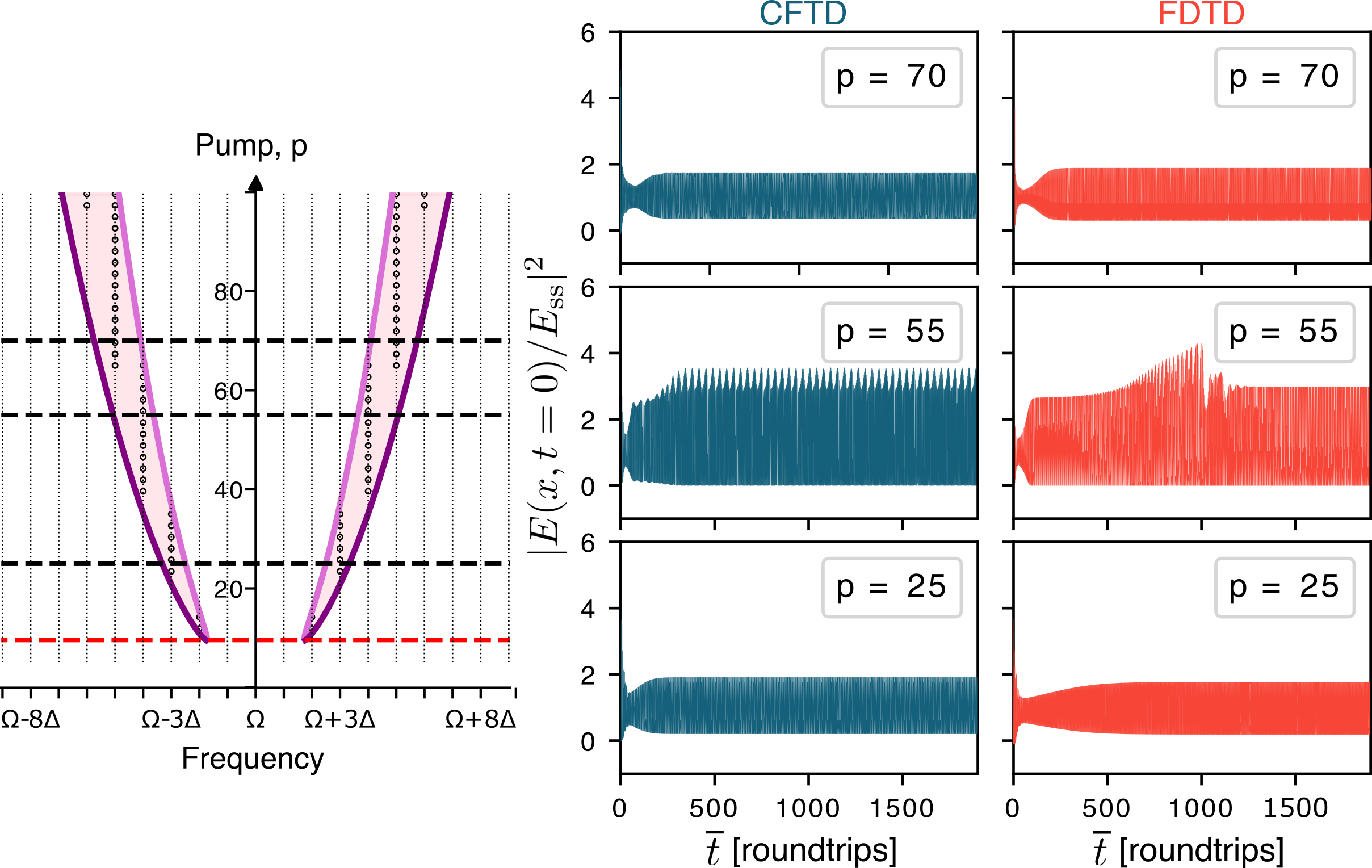}
    \caption[]{CFTD and FDTD comparison for decay parameters: $\tbar{\gamma}_{\perp}=5.0,\tbar{\gamma}_{\parallel}=0.5,\tbar{\kappa}=0.1$, and $n_{\rm R}=1.96$. Left panel: ST LSA where red dashed line marks the instability threshold and black dashed lines mark pump powers where time traces are shown. Regions marked by black open circles show the results of the modal LSA.  Right panel: Side by side comparison of $|E(x=0,t)|^2$ scaled by $\ESS$ for the first 2000 roundtrips of the CFTD (blue) and FDTD (red) results.} 
    \label{fig:above_thresh_g5}
\end{figure}

We consider here an additional intermediate ratio ${\tbar{\gamma}_{\perp}}/{\tbar{\gamma}_{\parallel}}=10$ as shown in Fig.~\ref{fig:above_thresh_g5}. The decay parameters are: $\tbar{\gamma}_{\perp}=5,\tbar{\gamma}_{\parallel}=0.5$ and $\tbar{\kappa}=0.1$. The ST LSA reveals that the first mode pair to become unstable is at the frequencies $\Omega \pm 2\Delta$. As we have previously seen, we can extract the unstable modes by looking at the intersection of each pump power and the instability band shaded in pink. The black circles in the same figure mark the modal linear stability results. The two analyses begin to diverge here for higher pump power as the modal approach seems to predict a smaller range of instability than the RNGH analysis. 

In this case, we show numerical CFTD and FDTD simulations for pump powers $p=25$ and $p=70$. In these cases only a single mode becomes unstable at frequencies $\Omega \pm 3\Delta$ and $\Omega \pm 5\Delta$ respectively. The time traces of the electric field computed using both CFTD and FDTD simulations show sustained oscillations in these cases, while the frequency spectra of the fields will reveal prominent peaks at the unstable mode pairs (not shown). However, this qualitative agreement is accompanied by certain discrepancies, such as the rate at which solutions reach steady state, and the steady-state oscillation amplitude.

The intermediate pump power result, $p=55$, describes a special case: here, \textit{two} mode pairs are simultaneously predicted to be unstable by the ST LSA ($\Omega \pm 4\Delta$ and $\Omega \pm 5\Delta$ fall simultaneously in the instability band). As seen in Fig.~\ref{fig:mainfig}, for simpler instabilities with a single unstable seed mode pair, this mode number determines the FSR spacing of the emergent multimode comb. With two adjacent unstable mode pairs, the nature of the comb waveform is not \textit{a priori} obvious, as the mode numbers are incompatible with a fixed FSR comb. The time domain electric field traces shown in Fig.~\ref{fig:above_thresh_g5} using both simulation methods reveal that the dynamics is more complex than the single mode pair case. 



\subsection{The role of the retained number of modes}

\begin{figure}[H]
    \centering
    \includegraphics[scale=0.8]{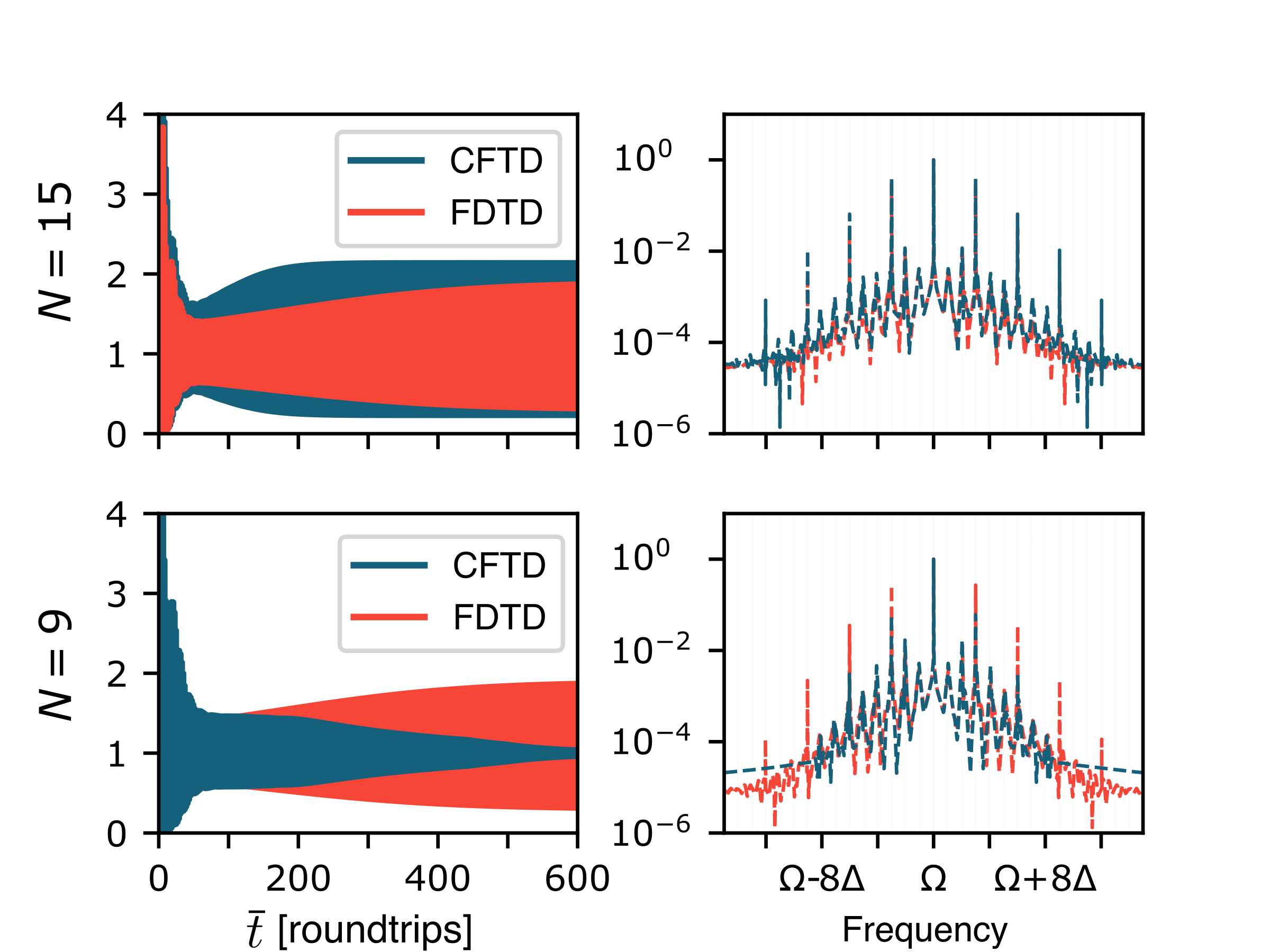}
    \caption[]{Comparison of dynamics of the scaled electric field intensity $|E(x=0,t)|^2$ and its FFT using CFTD (blue) and FDTD (red), above the instability threshold ($p = 25$) for two different numbers of modes included in the CFTD simulation. Decay parameters are $\tbar{\gamma}_{\perp}=5.0,\tbar{\gamma}_{\parallel}=0.5,\tbar{\kappa}=0.1$, and $n_{\rm R}=1.96$. }
    \label{fig:Mode_number}
\end{figure}

As a modal approach, one of the main parameters in the CFTD simulations is the number of modes. Generally, a larger number of modes leads to greater accuracy as defined by the discrepancy with the FDTD solution; however, if the number of modes is too large, it will lead to a greater simulation time without any benefit in accuracy and may even give rise to artifacts. In our analysis in the main text, the choice of number of modes has been guided by two factors. First, in order to produce an initial pulse that closely resembles a Gaussian pulse, at least 11 modes have been included in each simulation regardless of the relevant number of modes in the system. For some of the regimes we have considered, this is a sufficient number of modes. The second factor we have consulted is the LSA by including at least enough modes to cover the instability bands. To highlight the importance of the number of modes, we have shown in Fig.~\ref{fig:Mode_number} a comparison of two simulations where the difference lies only in the total number of modes included. The top panel shows the CFTD and FDTD time traces as well as spectra for a greater number of modes, in this case 15. We do see a discrepancy between the two methods in the time domain, where the FDTD is slower to reach the steady state oscillations. As we have analyzed in the main text, this parameter space belongs to a relatively complex regime, where the agreement between the two methods is more qualitative. However, the agreement over the modes present in each simulation is very good. In the bottom panel we show the same comparison between the two methods but now the number of modes included in the CFTD is only 9. The results are now qualitatively very different between the CFTD and FDTD where the CFTD settles to steady state and thus shows fewer modes to be present. In fact, the few sidebands that are present in the CFTD spectrum are due to the transient dynamics in the beginning of the time trace.

\nocite{*}


\end{document}